\documentclass{article}

\usepackage[preprint]{neurips_data_2024}

\usepackage[utf8]{inputenc} 
\usepackage[T1]{fontenc}    
\usepackage{hyperref}       
\usepackage{url}            
\usepackage{booktabs}       
\usepackage{amsfonts}       
\usepackage{nicefrac}       
\usepackage{microtype}      
\usepackage{xcolor}         
\usepackage{multirow}
\usepackage{tabularx}
\usepackage{adjustbox}
\usepackage{graphicx}
\usepackage{subcaption}
\usepackage{amsmath}
\usepackage{hyperref}

\title{From Theory to Therapy: Reframing SBDD Model Evaluation via Practical Metrics}

\author{
Bowen Gao\textsuperscript{1}\thanks{Equal contribution}\;\,, 
Haichuan Tan\textsuperscript{1,2}\footnotemark[1]\;\,, 
Yanwen Huang\textsuperscript{3}, 
Minsi Ren\textsuperscript{4}, 
Xiao Huang\textsuperscript{5}, \\
\textbf{Wei-Ying Ma\textsuperscript{1}},
\textbf{Ya-Qin Zhang\textsuperscript{1}}, 
\textbf{Yanyan Lan\textsuperscript{1}}\thanks{Correspondence to \texttt{lanyanyan@air.tsinghua.edu.cn}}  \\
\\
\textsuperscript{1}Institute for AI Industry Research (AIR), Tsinghua University \\
\textsuperscript{2}Department of Computer Science and Technology, Tsinghua University \\
\textsuperscript{3}Department of Pharmaceutical Science, Peking University \\
\textsuperscript{4}Institute of Automation, Chinese Academy of Sciences \\
\textsuperscript{5}College of Intelligence and Computing, Tianjin University \\
}

\begin{document}

\maketitle

\begin{abstract}
Recent advancements in structure-based drug design (SBDD) have significantly enhanced the efficiency and precision of drug discovery by generating molecules tailored to bind specific protein pockets. Despite these technological strides, their practical application in real-world drug development remains challenging due to the complexities of synthesizing and testing these molecules. The reliability of the Vina docking score, the current standard for assessing binding abilities, is increasingly questioned due to its susceptibility to overfitting. To address these limitations, we propose a comprehensive evaluation framework that includes assessing the similarity of generated molecules to known active compounds, introducing a virtual screening-based metric for practical deployment capabilities, and re-evaluating binding affinity more rigorously. Our experiments reveal that while current SBDD models achieve high Vina scores, they fall short in practical usability metrics, highlighting a significant gap between theoretical predictions and real-world applicability. Our proposed metrics and dataset aim to bridge this gap, enhancing the practical applicability of future SBDD models and aligning them more closely with the needs of pharmaceutical research and development.

\end{abstract}

\section{Introduction}

The field of Structure-Based Drug Design (SBDD) has experienced remarkable advancements in recent years, with the development of models such as Pocket2Mol \citep{peng2022pocket2mol}, TargetDiff \citep{guan20233d}, and MolCRAFT \citep{qu2024molcraft}. These models are at the forefront of enhancing the efficiency and precision of drug discovery by generating molecules designed to effectively bind to specific protein pockets. Despite these technological strides, the practical application of these models in real-world drug development remains a formidable challenge. The crux of this challenge lies in the verification of their efficacy, which is complicated by difficulties in synthesizing and testing these molecules in laboratory settings.

\begin{figure}[h]
\centering
\includegraphics[width=0.8\textwidth]{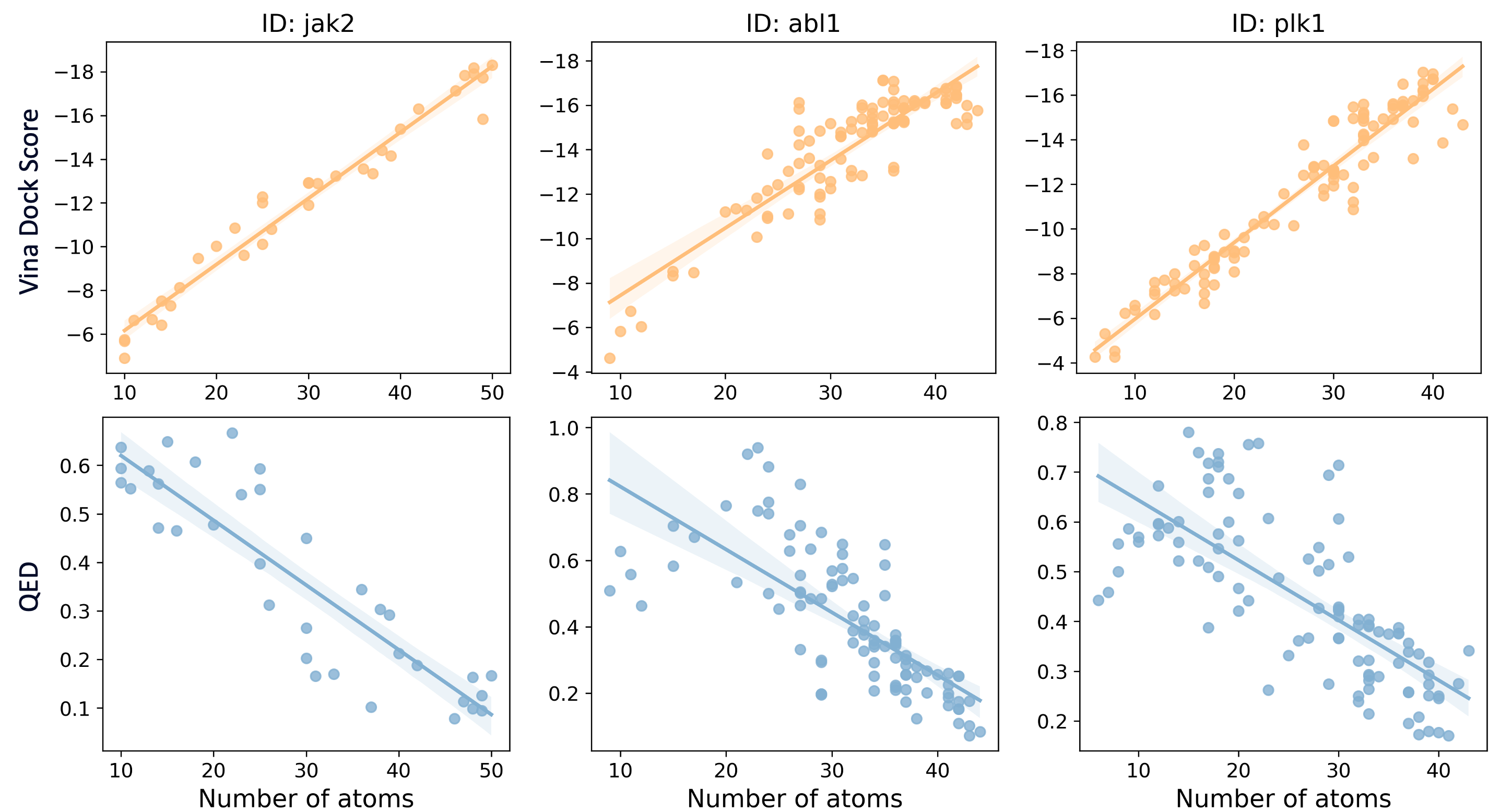} 
\caption{The relationship between Vina Dock Score/QED and number of atoms}
\label{fig:vina_qed} 
\vspace{-5pt}
\end{figure}

The Vina docking score \citep{eberhardt2021autodock,trott2010autodock} is currently the standard metric used to measure the binding abilities of molecules generated by SBDD models. It provides an estimate based on an empirical formula, serving as a proxy for binding affinity. Indeed, studies have shown that SBDD models can generate molecules with Vina docking scores that outperform those of reference ligands \citep{guan20233d,guan2024decompdiff,qu2024molcraft}, indicating a potentially huge improvement in this field. However, its reliability is increasingly being questioned. As shown in Figure \ref{fig:vina_qed}, Vina scores can be significantly inflated simply by increasing the number of atoms in a molecule, revealing a susceptibility to overfitting. This suggests that sole reliance on this metric may lead to overly optimistic evaluations of model performance. Moreover, as illustrated in \cite{gao2024rethinking}, recent developments in modeling have significantly improved Vina docking scores. However, the more critical estimation of specific binding ability, represented by the delta score, has remained unchanged or even deteriorated, and still falls short of the performance of reference ligands.


Furthermore, the practical synthesis of molecules generated by current SBDD models often proves to be complex and unfeasible, which significantly impedes their validation in wet-lab experiments \citep{bradshaw2019model, gao2020synthesizability}. This challenge, compounded by the long-term reliance on flawed metrics such as the Vina docking score, has led to a notable shortcoming in SBDD—a disconnect from practical applications. This gap is evident as the outputs of current SBDD models, while theoretically promising, prove challenging to utilize effectively in real-world settings. 

To address these critical shortcomings in the field of SBDD, we propose a comprehensive evaluation framework that extends beyond traditional estimation-based metrics. Recent attempts to incorporate SBDD-generated molecules into practical drug discovery processes have involved modifying these molecules into existing or more easily synthesizable forms \citep{moret2023leveraging}. Others have used these molecules as reference templates for virtual screening \citep{shen2024pocket}. Inspired by these approaches, our proposed evaluation framework includes several new metrics designed to directly assess the practical usability and deployment capabilities of generated molecules from different SBDD models.

This multifaceted framework assesses three levels of molecule evaluation: First, it evaluates the similarity of generated molecules to known active compounds and FDA-approved drugs, gauging their potential to be modified into viable drug candidates. Second, it introduces a virtual screening-based metric that directly measures the practical deployment capabilities of these molecules. Third, it continues to consider the estimated binding affinity, albeit in a more nuanced and critically evaluated manner.

By successfully meeting these comprehensive metrics, an SBDD model demonstrates a higher probability of producing molecules that are not only theoretically effective but also practically viable in real-world drug discovery settings. This approach aims to bridge the significant gap between the theoretical models of SBDD and their actual application in the pharmaceutical industry, paving the way for more reliable and efficacious drug development processes.

We conducted extensive experiments using our dataset, which includes data derived from real crystal structures and is divided based on local structural similarities of the pockets, to train and evaluate major SBDD models. Our results show that, from a practical deployment perspective, the molecules generated by current models fall significantly short of matching the quality of reference ligands. Despite achieving high Vina scores, their practical usability metrics reveal a substantial gap. Our proposed evaluation pipeline is designed to help bridge this gap, offering a direction that could enhance the practical applicability of future SBDD models.


\section{Related Work}

Structure-Based Drug Design involves generating small molecules with potential biological activity for a given protein pocket \citep{zhang2023systematic}. Traditionally, biological activity is measured using AutoDock Vina \citep{trott2010autodock}, a widely used docking software designed to predict the preferred binding orientation of a small molecule (ligand) when bound to a larger protein (receptor) target. 

Recently, with the development of deep generative models, several representative models have emerged. These include autoregressive models like AR \citep{luo20213d} and Pocket2Mol \citep{peng2022pocket2mol}, diffusion-based models like Targetdiff \citep{guan20233d}, and the newly developed Bayesian Flow Network model, MolCRAFT \citep{qu2024molcraft}.

Generally, these models are trained and tested on the CrossDocked dataset \citep{francoeur2020three}, which is created by cross-docking protein-small molecule pairs within the PDBbind dataset.

\section{Methods} \label{sec: methods}

\subsection{Evaluation Metrics that meet practical needs}

The ultimate goal of deep learning-based generative models in drug design is to generate drug-like molecules that can specifically bind to their intended targets. While recent advancements in SBDD models have shown impressive benchmark performance, even surpassing reference ligands in Vina docking scores, their practical application in pharmaceutical settings remains limited.

A significant limitation of current evaluation metrics is their inability to accurately gauge a model’s effectiveness in generating useful molecules. Traditionally, models have depended on Vina docking scores, which can be biased toward molecules that achieve high scores but are less effective in practical applications. For instance, as illustrated in Table \ref{tab:lbvs}, Vina scores are shown to be less predictive than other methods for virtual screening purposes. Additionally, it is demonstrated in Figure~\ref{fig:vina_qed} that merely increasing the molecular size can inflate Vina scores while simultaneously lowering the QED values, a measure of drug-likeness, thus exposing a vulnerability to overfitting. More importantly, despite improvements in Vina docking scores, recent research indicates that the delta score—a crucial metric for assessing specific binding capability—still significantly lags behind that of reference ligands. This discrepancy could misdirect model development into paths far removed from practical needs, highlighting the inadequacy of relying solely on Vina docking scores and underscoring the necessity for more relevant and accurate evaluation criteria.

Another challenge is the synthetic feasibility of molecules generated by deep learning models. Recent studies indicate these molecules are often difficult to synthesize \citep{bradshaw2019model, gao2020synthesizability}, hindering wet-lab validation and practical application.

To address the limitations of current metrics in SBDD, we propose a reevaluation of how model outputs are assessed. Recent research has involved manually modifying generated molecules into synthesizable structures for wet-lab validation, drawing on the expertise of medicinal chemists who typically adjust molecules based on actives targeting the same biological structures and approved drugs \citep{moret2023leveraging}. Building on this approach, we suggest shifting the evaluation paradigm. Rather than solely aiming for molecules ready for direct wet-lab experiments and binding affinity estimation, we should assess their potential to resemble active compounds or known drugs. This perspective focuses on the feasibility of transforming generated molecules into practically useful compounds. A similarity or distance-based metric could then be employed to gauge this practical utility, reflecting a more realistic and applicable measure of a molecule’s value in drug development.


Furthermore, leveraging the capability of generative models in virtual screening to serve as templates for identifying similar compounds has shown promise, achieving significant hit rates \citep{shen2024pocket}. Thus, we suggest adding a virtual screening metric to evaluate how well a generated molecule can discriminate between active and inactive compounds, providing a direct measure of its utility in drug discovery.

\begin{figure}[htbp]
\centering
\includegraphics[width=\textwidth]{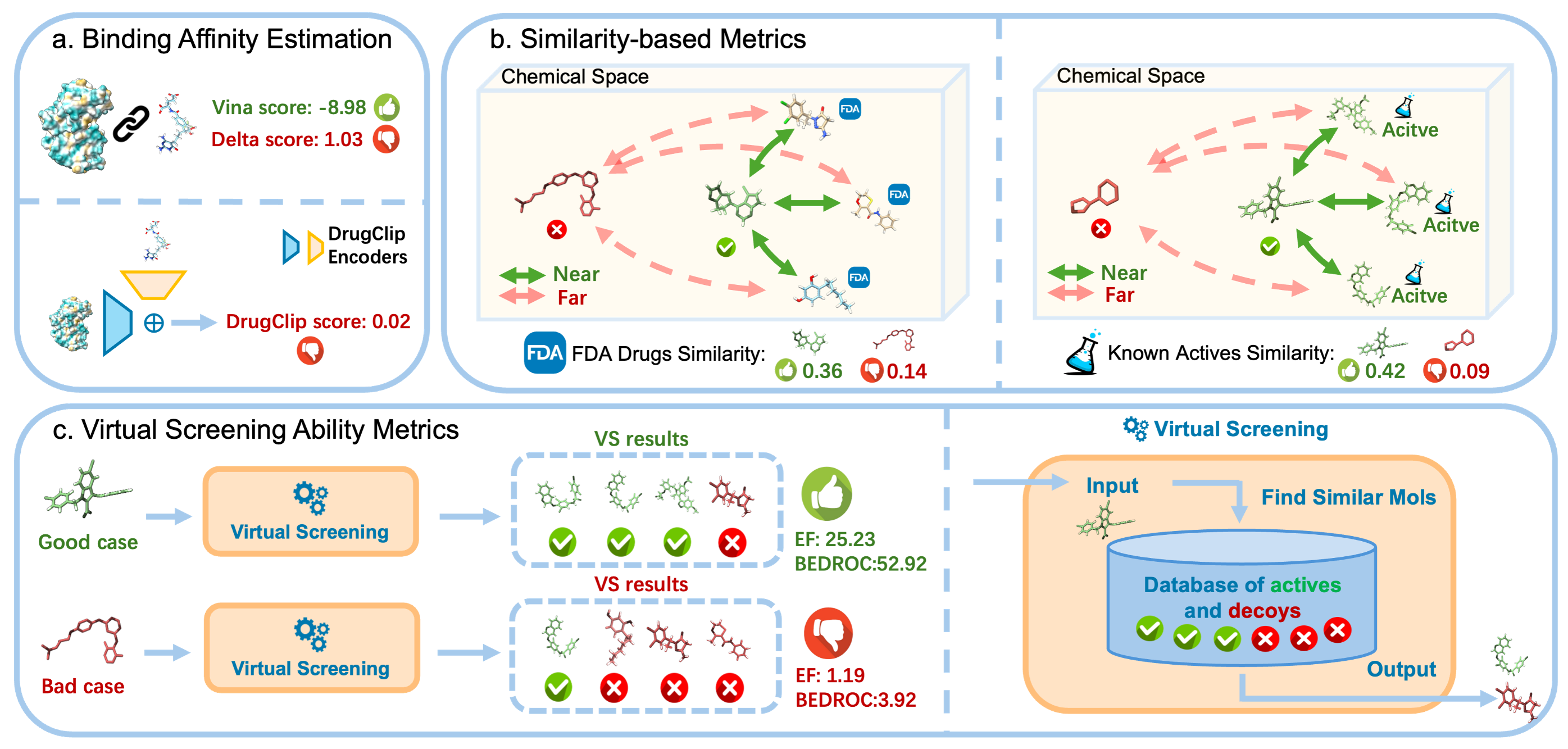} 
\caption{Our three-level evaluation metrics include: (a) Binding affinity estimation, which encompasses the Vina docking score, delta score, and DrugCLIP score; (b) Similarity-based metrics that assess the resemblance between generated molecules and known actives, as well as drugs on the FDA-approved list; (c) Virtual screening ability metrics that evaluate the capability of the generated molecules to differentiate between actives and decoys when used as reference templates.}
\label{fig:metrics} 
\end{figure}

In summary, as shown in Figure \ref{fig:metrics}, our proposed evaluation metrics for assessing the effectiveness of deep learning-based generative models in drug design are structured across three levels:

\begin{enumerate}
    \item \textbf{Binding Affinity Estimation:} This metric directly measures the binding capabilities of generated molecules to target structures, providing a direct assessment of their potential therapeutic effectiveness.
    \item \textbf{Similarity-Based Metrics:} These metrics evaluate the potential for modification and optimization of generated molecules. They assess how closely these molecules resemble known active compounds, facilitating easier synthesis and optimization.
    \item \textbf{Virtual Screening-Based Metrics:} These metrics determine the ability of generated molecules to distinguish between active and inactive compounds. This is crucial for identifying potential drug candidates that are more likely to succeed in later stages of drug development.
\end{enumerate}

We introduce the details of those metrics.

\paragraph{Binding Affinity Estimation}

The most commonly used binding affinity estimation methods are docking scores generated by docking software like Vina, as utilized in previous works \citep{trott2010autodock}. In this benchmark, we continue to use docking scores as one of the metrics to reflect the binding estimation. In addition to the conventional docking score, we also use the delta score proposed by \citep{gao2024rethinking}, which provides a good estimation of the specific binding ability of generated molecules.

Beyond docking scores, we also employ a machine learning-based scoring function. DrugCLIP \cite{gao2023drugclip} has demonstrated outstanding performance in virtual screening, making it a valuable evaluation metric for assessing the binding potential of generated molecules. Here, we denote the ligand and pocket as $l$ and $p$, respectively. The ligand encoder and pocket encoder pretrained by DrugCLIP are denoted as $\sigma_{\text{ligand}}$ and $\sigma_{\text{pocket}}$. The DrugCLIP score is then defined as the dot product of the encoded representations of the ligand and pocket:

\begin{equation}
\text{DrugCLIP Score} (l) = \sigma_{\text{ligand}}(l) \cdot \sigma_{\text{pocket}}(p).
\end{equation}

\paragraph{Similarity to Known Drugs and Actives}

Our framework maps molecules into a feature space using molecular fingerprints, specifically 2D Extended Connectivity Fingerprints (ECFP) and 3D Extended Three-dimensional Fingerprints (E3FP), which are 1024-dimensional bit vectors. To enhance this, we integrate deep learning-based encoders like Uni-Mol \cite{zhou2022uni} and the DrugCLIP \citep{gao2023drugclip} molecular encoder, which aligns with binding pockets.

Given a target with \( N \) known actives \( a_1, a_2, \ldots, a_N \), the Active Similarity score is:

\[
\text{Active Similarity} (l) = \max_{i \in \{1, 2, \ldots, N\}} \left( \sigma(l) \cdot \sigma(a_i) \right).
\]

We also assess drug-likeness by comparing the similarity of generated molecules to \( 2582 \) FDA-approved drugs:

\[
\text{FDA Similarity} (l) = \max_{j \in \{1, 2, \ldots, 2582\}} \left( \sigma(l) \cdot \sigma(d_j) \right).
\]

where \( d_j \) represents the \( j \)-th molecule in the FDA-approved drug database.

\paragraph{Virtual Screening Ability}

We use the molecules generated by our models as references to retrieve similar compounds from the compound library, evaluating the accuracy of identifying known actives. The library comprises experimentally validated actives and decoys. Decoys are structurally analogous to actives but have been experimentally confirmed to be incapable of binding, rendering the task of distinguishing actives from decoys particularly arduous and complex. Similarly, we can utilize a variety of different encoders for virtual screening.

Specifically, we use the BEDROC and EF metrics to evaluate the effectiveness of virtual screening. BEDROC incorporates exponential weights that assign greater importance to early rankings. In the context of virtual screening, the commonly used variant is BEDROC$_{85}$, where the top 2\% of ranked candidates contribute to 80\% of the BEDROC score. The formal definition is shown in equation \ref{eq:bedroc}, where $\mathrm{NTB}_\alpha$ is the number of true binders in the top $\alpha \%$. Enrichment Factor (EF) is also a widely used metric, calculated as $\mathrm{EF}_\alpha=\frac{\mathrm{NTB}_\alpha}{\mathrm{NTB}_t \times \alpha}$, where $\mathrm{NTB}_t$ is the total number of binders in the entire screening pool. Table \ref{tab:lbvs} demonstrates that using real ligands as templates for virtual screening yields better results compared to virtual screening with docking software, validating the reliability of our proposed metric. 

\begin{equation}
\operatorname{BEDROC}_\alpha=\frac{\sum_{i=1}^{\mathrm{NTB}_t} e^{-\alpha r_i / N}}{R_\alpha\left(\frac{1-e^{-\alpha}}{e^{\alpha / N}-1}\right)} \times \frac{R_\alpha \sinh (\alpha / 2)}{\cosh (\alpha / 2)-\cosh \left(\alpha / 2-\alpha R_\alpha\right)}+\frac{1}{1-e^{\alpha\left(1-R_\alpha\right)}}.
\label{eq:bedroc}
\end{equation}


\vspace{-2pt}
\begin{table}[h]
\caption{Results from virtual screening using docking software, and using real ligands as templates for similarity searches. Similarity is determined through various molecular fingerprints and deep learning encoders.}
\label{tab:lbvs}
\centering
\begin{adjustbox}{width=0.5\textwidth}
\begin{tabular}{cc|cc}
\toprule
 &    & BEDROC & EF@1  \\ \hline
                                   \noalign{\vskip 2pt} 
\multirow{2}{*}{Docking} 
& Glide    & 40.7   & 16.18 \\
& Vina     &   -     & 7.32  \\ \hline
\noalign{\vskip 2pt} 
\noalign{\vskip 2pt} 
\multirow{2}{*}{Fingerprints}      

& 2D   & 39.32  & 24.95  \\
\noalign{\vskip 2pt} 
& 3D  & 23.77  & 14.30 \\ \hline
\noalign{\vskip 2pt}                             
\multirow{2}{*}{Encoder}      
& UniMol   & 13.39  & 7.48  \\
\noalign{\vskip 2pt} 
& DrugCLIP        & 45.43  & 29.23 \\ 
\noalign{\vskip 2pt} 
\bottomrule
\end{tabular}
\end{adjustbox}
\end{table}
\vspace{-10pt}

\subsection{Test Dataset}

Previous work primarily tested models on the CrossDocked \citep{francoeur2020three} test set, which has several limitations. It is randomly selected, lacks diversity checks, and is derived from synthesized data rather than real-world crystal structures. Additionally, each pocket in CrossDocked is paired with only one ground truth ligand, even though a single pocket can bind to multiple different ligands.

To address these limitations, we propose a new test set for SBDD based on the well-known virtual screening benchmark DUD-E \citep{mysinger2012directory} and LIT-PCBA \citep{tran2020lit}. Our test set comprises 101 targets filtered from DUD-E and 15 targets from LIT-PCBA, with accurately recorded ligand and protein files, and includes a diverse range of protein types as shown in Appendix in Supplementary. Each target is supplemented with a significant number of actives and decoys, facilitating the evaluation of similarity and distance-based scores as well as virtual screening performance. A robustly generated molecule should effectively distinguish between actives and decoys. On average, each target in our test set contains 224.4 actives and 50 decoys for each active. This comprehensive and realistic benchmark ensures that SBDD models are evaluated against diverse and authentic data.

\subsection{Training and Validation Dataset}




Existing Models are commonly trained and tested on the CrossDocked \citep{francoeur2020three} dataset, produced by docking software, which consists of synthesized protein-ligand complexes. In contrast, our benchmark utilizes real protein-ligand conformations from experimental crystal structures in PDBbind \citep{wang2004pdbbind}. Actually, ligands in PDBbind has a higher docking score as well as delta score for specific binding ability. Details shown in Appendix \ref{sec: appendix compare}

We refined the PDBbind dataset by excluding complexes with nuclear attachment and inaccurately recorded ligands. Then split into a 9:1 training and validation set. To assess the SBDD model’s generalization across diverse pockets, we removed samples with similar pockets using FLAPP for pocket alignment and similarity assessment. We remove all pockets from the training set with align rate more than 0.6 or 0.9 to the test set pocket. After the removal, the 0.6 version has 12344 pairs remaining while the 0.9 version has 17519 pairs remaining. Details can be found in the Appendix attached in the supplementary.

\section{Experiments}

\subsection{Tested Models}

\vspace{-5pt}
We select representative deep learning-based models for structure-based drug design evaluation. For voxel-grid based model, we use LiGAN \citep{ragoza2022chemsci}. For autoregressive models, we choose AR \citep{luo20213d} and Pocket2Mol \citep{peng2022pocket2mol}. For diffusion models, we select Targetdiff \citep{guan20233d}. Additionally, MolCRAFT \citep{qu2024molcraft} is included as a generative model based on a Bayesian flow network.
\vspace{-5pt}

\subsection{Results of Binding Ability Estimation}

\begin{table}[ht]
\centering
\caption{Evaluation Results for binding ability estimation. Results for Reference Ligands and best results are shown in \textbf{bold text}.}
\label{tab:binding ability}
\begin{adjustbox}{width=0.9\textwidth}
\begin{tabular}{cc|ccc}
\toprule
&    & Vina Docking Score $\downarrow$    & Delta Score $\uparrow$   & DrugCLIP score $\uparrow$      \\ \hline
\noalign{\vskip 2pt} 
\multicolumn{2}{c|}{Reference Ligand}        & \textbf{-9.363}     & \textbf{2.686}   & \textbf{0.508}     \\ \hline
\noalign{\vskip 2pt} 
\multirow{5}{*}{\begin{tabular}[c]{@{}c@{}}PDBbind \\ 60\end{tabular}} 
& LiGAN       & -5.175    & 0.037  & -0.016      \\
 & AR         & -7.255   & 0.483  & 0.009    \\
 & Pocket2Mol & -7.640    & 0.531  & -0.005   \\
& TargetDiff  & -9.562    & 0.325  & 0.099 \\
& MolCRAFT    & \textbf{-9.788}    & \textbf{0.973}  & \textbf{0.145}     \\ \hline
\noalign{\vskip 2pt} 
\multirow{5}{*}{\begin{tabular}[c]{@{}c@{}}PDBbind \\ 90\end{tabular}} 
& LiGAN       & -6.577    & 0.107  & -0.000     \\
 & AR         & -7.340   & 0.523  & 0.007      \\
 & Pocket2Mol & -8.195    & 0.599  & -0.006     \\
& TargetDiff  & -9.711    & 0.238  & 0.095      \\
& MolCRAFT & \textbf{-9.778}    & \textbf{1.163}  & \textbf{0.173}     \\ 
\bottomrule
\end{tabular}
\end{adjustbox}
\end{table}

In Table \ref{tab:binding ability}, it is evident that both TargetDiff and MolCRAFT outperform the reference ligand in terms of average docking scores. However, when considering the Delta Score and DrugCLIP score, they lag significantly behind the reference ligand. Specifically, MolCRAFT, while the best performer among the evaluated methods, still shows a considerable disparity in Delta Score (0.973 vs. 2.686) and DrugCLIP score (0.173 vs. 0.508) when compared to the reference ligand. It is noteworthy that although TargetDiff achieves a competitive docking score, its Delta Score is inferior to those of autoregressive-based methods. This suggests that TargetDiff’s high docking score might be the result of overfitting large atom numbers. 

\subsection{Results of Similarity-Based Metrics}

\vspace{-10pt}
\begin{table}[ht!]
\centering
\caption{Evaluation Results for Similarity-Based Metircs. Results for Reference Ligands and best results are shown in \textbf{bold text}.}
\label{tab:similarity}
\begin{adjustbox}{width=1\textwidth}
\begin{tabular}{cc|cc|cc|cc|cc}
\toprule
&  & \multicolumn{2}{c|}{2D Fingerprints} & \multicolumn{2}{c|}{3D Fingerprints}  & \multicolumn{2}{c|}{UniMol} & \multicolumn{2}{c}{DrugCLIP}\\
&  & FDA     & Active    & FDA     & Active  & FDA     & Active & FDA     & Active  \\ \hline
\noalign{\vskip 2pt}
\multicolumn{2}{c|}{Reference Ligand}        & \textbf{0.348}   & \textbf{0.588}   & \textbf{0.139}   & \textbf{0.230} & \textbf{0.975} & \textbf{0.973} & \textbf{0.749} & \textbf{0.870} \\ \hline
\noalign{\vskip 2pt}
\multirow{5}{*}{\begin{tabular}[c]{@{}c@{}}PDBbind \\ 60\end{tabular}} 
& LiGAN       & 0.231 & 0.131  & 0.117 & 0.109 & 0.950 & 0.922 & \textbf{0.783} & 0.366 \\
& AR          & 0.237 & 0.157  & 0.116 & 0.122 & 0.966 & 0.949 & 0.747 & 0.449 \\
& Pocket2Mol  & \textbf{0.286} & 0.179  & \textbf{0.137} & 0.143 & 0.964 &0.946  & 0.735 & 0.414 \\
& TargetDiff  & 0.209 & 0.169  & 0.132 & 0.143 &0.967  & 0.958 & 0.683  & 0.478 \\
& MolCRAFT & 0.258  & \textbf{0.208} & 0.129 & \textbf{0.161} & \textbf{0.971} & \textbf{0.965} & 0.676 & \textbf{0.522} \\ \hline
\noalign{\vskip 2pt}
\multirow{5}{*}{\begin{tabular}[c]{@{}c@{}}PDBbind \\ 90\end{tabular}} 
& LiGAN       & 0.236 & 0.136  & 0.117 & 0.112 & 0.952 & 0.931 & \textbf{0.733} & 0.397 \\
& AR          & 0.229 & 0.161  & 0.113 & 0.125 & 0.965 & 0.951 & 0.730 & 0.468 \\
& Pocket2Mol  & \textbf{0.283} & 0.185  & \textbf{0.134} & 0.146 & 0.963 & 0.945 & 0.732 & 0.419 \\
& TargetDiff  & 0.208 & 0.169  & 0.132 & 0.142 & 0.967 & 0.958 & 0.686 & 0.477 \\
& MolCRAFT & 0.258 & \textbf{0.214}  & 0.128 & \textbf{0.163} & \textbf{0.972} & \textbf{0.966} & 0.681 & \textbf{0.547} \\
\bottomrule
\end{tabular}
\end{adjustbox}
\end{table}

The evaluation results using similarity-based metrics are presented in Table \ref{tab:similarity}. Pocket2Mol demonstrates the highest similarity to known drugs on the FDA-approved list based on 2D Fingerprints, while MolCRAFT excels in generating molecules that closely resemble known active compounds for specific targets. However, both models, along with other methods tested, significantly underperform compared to the reference ligand. A notable observation is that the reference ligand shows greater similarity to known active compounds than to FDA-approved drugs in general. In contrast, the generative models display higher similarity to the broader category of FDA-approved drugs rather than to specific known actives. This suggests that current structure-based drug design (SBDD) models still lack effective conditional generation capabilities, highlighting a key area for further development. The distribution plots for similarity-based metrics across all targets in DUD-E are shown in Appendix \ref{sec: appendix fda sim} \ref{sec: appendix actives sim}.

\subsection{Results of Virtual Screening-Based Metircs}
\begin{table}[ht!]
\caption{Evaluation Results for Virtual Screening-Based Metircs. Results for Reference Ligands and best results are shown in \textbf{bold text}.}
\label{tab:vs}
\begin{adjustbox}{max width=1\textwidth}
\begin{tabular}{cc|cc|cc|cc|cc}
\toprule
&    & \multicolumn{2}{c|}{2D Fingerprints}   & \multicolumn{2}{c|}{3D Fingerprints}   & \multicolumn{2}{c|}{UniMol}   & \multicolumn{2}{c}{DrugCLIP}      \\
&    & BEDROC &EF & BEDROC & EF &BEDROC & EF &BEDROC & EF\\ \hline
\noalign{\vskip 2pt}
\multicolumn{2}{c|}{Reference Ligand}        & \textbf{39.32}   & \textbf{24.95}   & \textbf{23.77}   & \textbf{14.30}   & \textbf{13.39}   & \textbf{7.48}   & \textbf{45.43}   & \textbf{29.23}    \\ \hline
\noalign{\vskip 2pt}
\multirow{5}{*}{\begin{tabular}[c]{@{}c@{}}PDBbind \\ 60\end{tabular}} 
& LiGAN       & 2.306   & 1.079   & 2.216   & 1.007   & 2.024   & 0.837   & 1.581   & 0.655     \\
& AR          & 4.938   & 2.567   & 4.407   & 2.196   & 3.266   & 1.501   & 3.698   & 1.796     \\
& Pocket2Mol  & 5.976   & 3.054   & 4.192   & 2.047   & 2.335   & 1.033   & 3.667   & 1.827     \\
& TargetDiff  & 4.062 & 1.957 & 3.747 & 1.763 & 2.995 & 1.340 & 4.400 & 2.260     \\
& MolCRAFT & \textbf{7.584}   & \textbf{3.953}   & \textbf{5.521}   & \textbf{2.792}   & \textbf{4.868}   & \textbf{2.357}   & \textbf{7.265}   & \textbf{3.968}     \\ \hline
\noalign{\vskip 2pt}
\multirow{5}{*}{\begin{tabular}[c]{@{}c@{}}PDBbind \\ 90\end{tabular}} 
& LiGAN       & 2.240   & 1.004   & 2.610   & 1.183   & 2.151   & 0.932   & 1.450   & 0.636    \\
& AR          & 4.946   & 2.522   & 4.156   & 2.053   &  3.115  & 1.384    &  4.223  & 2.139     \\
& Pocket2Mol  & 6.277   & 3.240   & 4.359  & 2.174     & 2.970   & 1.312   & 3.215   & 1.541     \\
& TargetDiff  &4.431   & 2.174   & 3.888   & 1.861   & 3.967   & 1.828   & 4.122   & 2.117     \\
& MolCRAFT & \textbf{9.032}   & \textbf{4.825}   & \textbf{6.423}   & \textbf{3.284}   & \textbf{5.299}   & \textbf{2.456}   & \textbf{9.782}   & \textbf{5.549}     \\ 
\bottomrule
\end{tabular}
\end{adjustbox}
\end{table}

The results for virtual screening ability metrics are presented in Table \ref{tab:vs}. MolCRAFT outshines all other methods, regardless of the feature extractor utilized. Nonetheless, it underperforms when benchmarked against reference ligands. Optimistically, using the DrugCLIP molecular encoder for encoding generated molecules for similarity-based virtual screening enables MolCRAFT to achieve a virtual screening efficacy comparable to that of Vina, with an enrichment factor of 5.549 versus Vina’s 7.32 (Table ~\ref{tab:lbvs}). This result is particularly encouraging as the speed of virtual screening with generated molecules is significantly faster than using docking software to dock all candidate compounds. The distribution plots for the enrichment factor of all targets in DUD-E across various models are displayed in \ref{fig:ef_dist} and detailed in Appendix \ref{sec:appendix vs}.

\subsection{Results of LIT-PCBA dataset}

\begin{table}[ht!]
\centering
\caption{Evaluation Results for Similarity-Based Metrics on LIT-PCBA dataset. Results for Reference Ligands and best results are shown in \textbf{bold text}.}
\label{tab:lit pcba}
\begin{adjustbox}{width=0.8\textwidth}
\begin{tabular}{c|cc|cc}
\toprule
 & \multicolumn{2}{c|}{Similarity to Actives} & \multicolumn{2}{c}{Virtual Screening} \\
 \noalign{\vskip 4pt}
 & Fingerprints & DrugCLIP& BEDROC & EF \\ 
\midrule
Reference Ligand & \textbf{0.269} & \textbf{0.613} & \textbf{4.332} & \textbf{3.641} \\ 
\noalign{\vskip 2pt}
\hline
\noalign{\vskip 2pt}
LiGAN & 0.141 & 0.529  & 1.527 & 0.889 \\
AR & 0.157  & \textbf{0.558} & 1.879 & 1.361 \\
Pocket2Mol & 0.187  & 0.538 & 2.363 & \textbf{1.711} \\
TargetDiff & 0.167  & 0.508 & 2.095 & 1.237 \\
MolCRAFT & \textbf{0.189}  & 0.556 & \textbf{2.498} & 1.577 \\ 
\bottomrule
\end{tabular}
\end{adjustbox}
\end{table}

In addition to DUD-E, which utilizes decoys designed according to specific rules, we also benchmark the models against LIT-PCBA, an alternative virtual screening dataset. This dataset consists of decoys that have been actualized in wet-lab experiments and exhibit minimal bioactivity. The threshold used to differentiate between active compounds and decoys in LIT-PCBA is more relaxed, making it a significantly more stringent test.

Table \ref{tab:lit pcba} shows that LIT-PCBA is indeed a difficult dataset, with an enrichment factor of 3.64 when using actual ligands as a reference, indicating a modest improvement over random selection. MolCRAFT remains the top-performing model overall.

\subsection{Analysis}

\begin{figure}[ht!]
    \centering
    \begin{subfigure}[b]{0.43\textwidth}
        \centering
        \includegraphics[width=\textwidth]{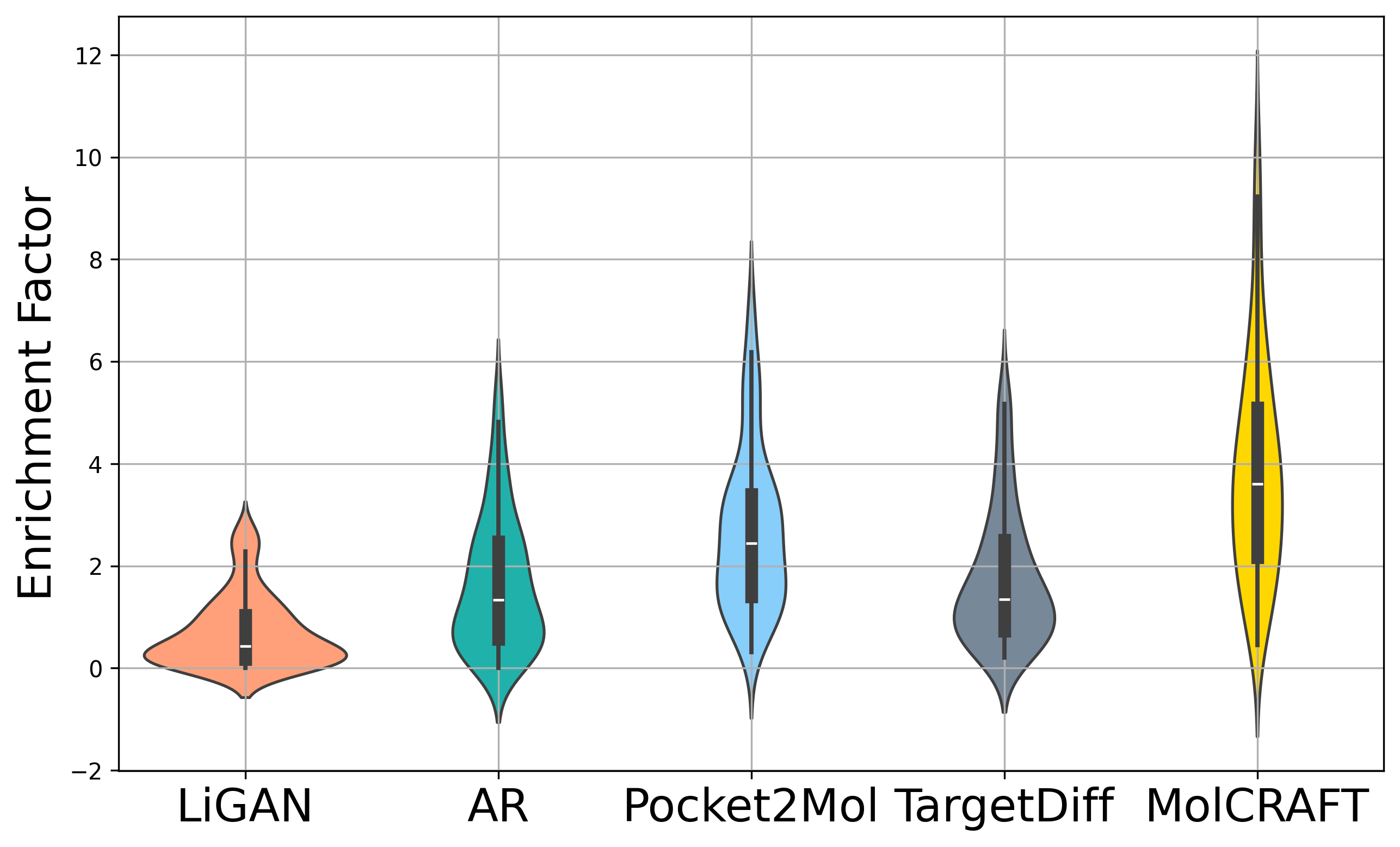}
        \caption{}
        \label{fig:ef_dist}
    \end{subfigure}
    \hfill 
    \begin{subfigure}[b]{0.56\textwidth}
        \centering
        \includegraphics[width=\textwidth]{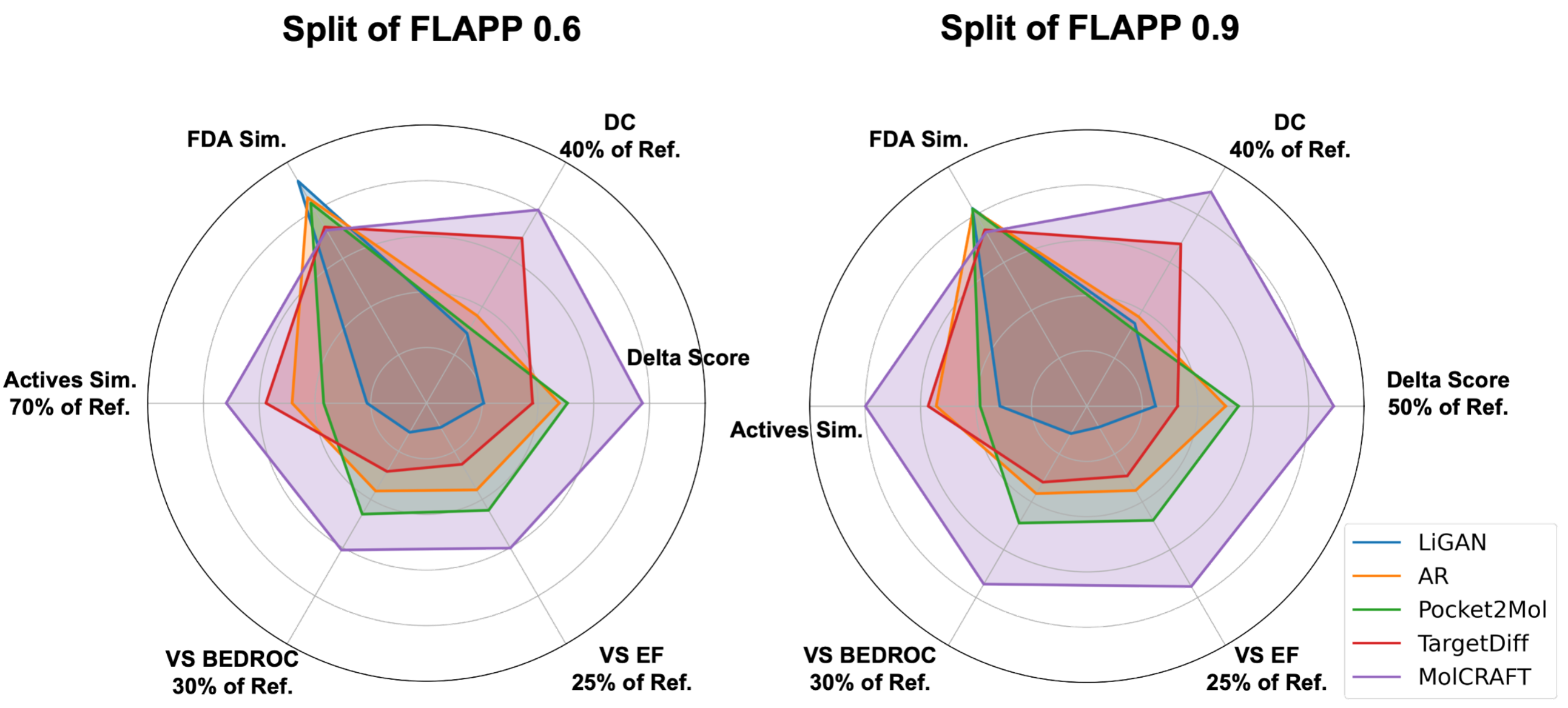}
        \caption{}
        \label{fig:rader}
    \end{subfigure}
    \caption{(a) Distribution of Enrichment Factors of different models across different targets in DUD-E. (b) Radar plot that shows the performance of different methods on part of our multifaceted metrics.}
    \label{fig:case}
\vspace{-10pt}

\end{figure}

We highlight some interesting findings from our benchmark results:

\begin{itemize}
\item \textbf{TargetDiff}: While achieving commendable Vina docking scores, TargetDiff's overall performance is suboptimal. The generated molecules show poor delta scores, low similarity to known actives, and struggle to distinguish between active and decoy molecules, even compared to autoregressive methods. This suggests diffusion-based SBDD models \citep{guan2024decompdiff, huang2024proteinligand}, although being very popular, may be overfitting to Vina scores without generating truly useful molecular structures.

\item \textbf{Pocket2Mol}: Ranked second-best for similarity-based metrics and virtual screening using fingerprint-based extractors, its performance declines with deep encoder-based methods. The autoregressive nature limits its ability to generate larger molecules, creating a distribution gap. Despite this, its good performance on virtual screening indicates that Pocket2Mol can generate valuable substructures or functional groups, though smaller molecules impact its docking scores negatively.

\item \textbf{MolCRAFT}: As depicted in Figure \ref{fig:rader}, MolCRAFT ranks as a top performer across multiple metrics and excels in virtual screening. It achieves results comparable to Vina docking but with greater speed, demonstrating that generative models are an efficient and promising approach to drug discovery. This provides a practical alternative to testing deep learning-generated molecules directly in wet-lab experiments.

\item \textbf{General Observations}: Despite higher docking scores, current models still lag in other critical metrics such as delta score, similarity to known actives, and virtual screening ability compared to reference ligands. Relying primarily on Vina docking scores can give an overly optimistic view of SBDD models’ effectiveness, highlighting the need for a more holistic approach to performance assessment.
\end{itemize}

\vspace{-4pt}

\subsection{Importance of using similarity based metrics}

\vspace{-7pt}
\begin{figure}[ht!]
    \centering
    \includegraphics[width=1\textwidth]{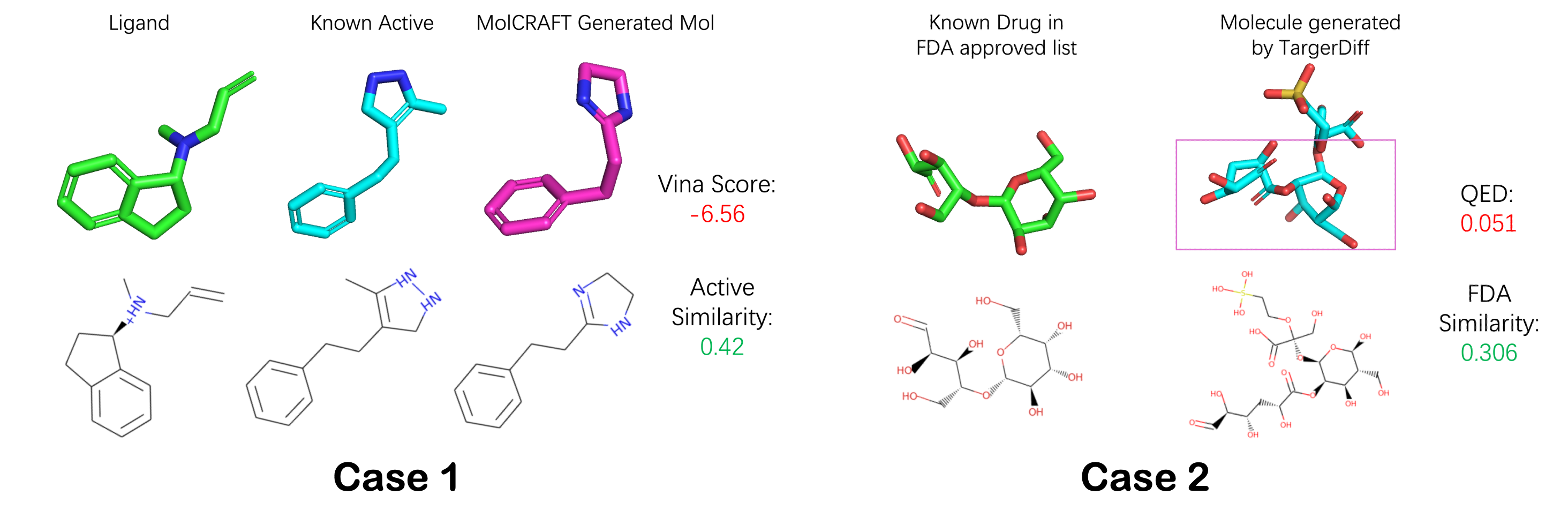}
    \caption{Cases that show the importance of using similarity-based metrics to evaluate the effectiveness of generated molecules}
    \label{fig:case all}
\end{figure}



Case 1 in Figure\ref{fig:case all} depicts a molecule generated by MolCRAFT that, although diverging from the reference ligand, aligns with a known active molecule of target AOFB, achieving a Morgan fingerprint similarity score of 0.42 but receiving a lower Vina score of -6.56. This example underscores the necessity of using more practical metrics, rather than solely relying on Vina scores, and emphasizes the importance of considering all known actives in addition to reference ligands for comprehensive model evaluation.

Case 2 in Figure \ref{fig:case all}, a molecule from the TargetDiff model exhibits significant similarity to a molecule on the FDA approved list, despite its low QED score. This suggests its potential as a drug candidate through targeted modifications. This case highlights the importance of our multifaceted metrics for a comprehensive assessment of drug-likeness in generated molecules.



\subsection{Conclusion}

In this paper, we conduct a comprehensive benchmark of current SBDD models. Unlike previous evaluations that primarily focused on Vina docking scores as a metric to estimate the binding ability of generated molecules, we propose evaluating them from a practical deployment perspective. This includes assessing the proximity of the generated molecules to known actives and drugs, as well as using these molecules as templates for virtual screening. Our results reveal that although previous methods demonstrate superior performance in terms of docking scores compared to reference ligands, they significantly lag behind in more practical metrics as we proposed. This discrepancy highlights the over-optimism of previous evaluations. We hope our proposed metrics will guide the development of SBDD models in a more reliable direction and bridge the gap between generative models and their practical applications.

\bibliography{sbdd}
\bibliographystyle{plainnat}

\newpage

\appendix

\section{Dataset and Code Availability}

The dataset we use for this benchmark is available at \\\href{https://huggingface.co/datasets/bgao95/Practical_SBDD}{https://huggingface.co/datasets/bgao95/Practical\_SBDD}

The code and instruction used to train the models for this benchmark is available at \\ \href{https://github.com/bowen-gao/sbdd_practical_evaluation}{https://github.com/bowen-gao/sbdd\_practical\_evaluation}

The dataset is hosted by Hugging Face. The license is CC BY 4.0. We bear all responsibility in case of violation of rights.

The data we are using/curating doesn't contain personally identifiable
information or offensive content.

\section{Training and Validation Set Details}

We constructed our training and validation sets using data from PDBbind, renowned for its highly reliable, experimentally observed structures. The dataset underwent rigorous filtering to ensure quality and relevance. Initially, we excluded all ligands that RDKit could not correctly interpret, including those with erroneous molecular structures and discontinuous molecules. Subsequently, we selected pockets from target PDB files using a 10 Å distance threshold from the ligand. We further refined the dataset by excluding pockets containing nucleic acids (DNA/RNA) and repairing non-standard residues within the pockets. Rare non-standard residues that could not be repaired were removed. Additionally, protein pockets with fewer than 100 atoms were discarded. This comprehensive filtering process yielded a final set of 19,438 protein pockets, which we then used to construct our training and validation datasets.

\begin{figure}[htbp]
\centering
\includegraphics[width=0.6\textwidth]{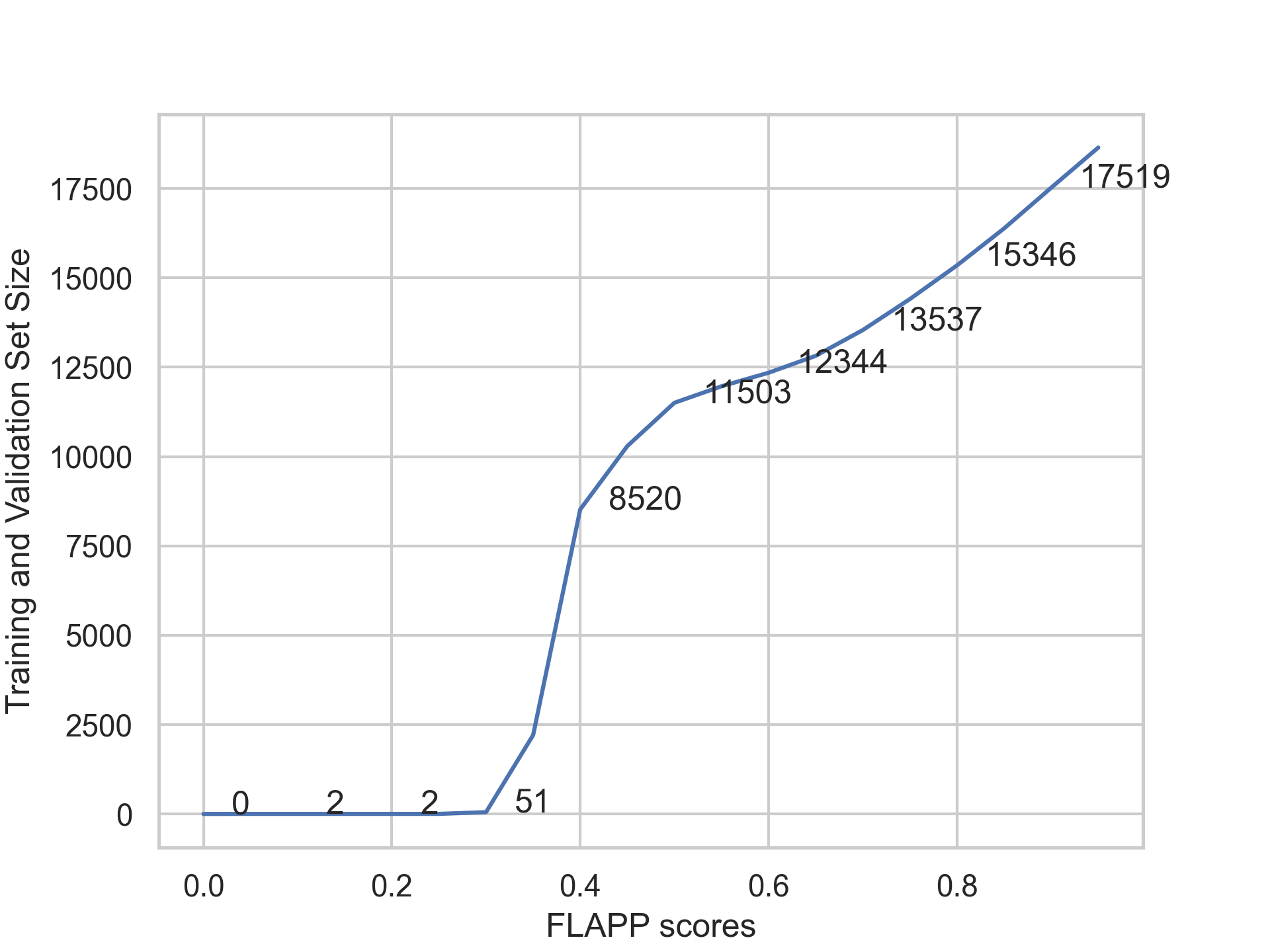}
\caption{Relationship between FLAPP Score Threshold and Dataset Sizes}
\label{fig:flapp_dist}
\end{figure}

To assess the ability of the SBDD generation model to generalize to novel pocket types, we implemented a homology reduction based on pocket structural similarity between the training and test sets. Utilizing FLAPP, we aligned pockets from the training set with those from the test set, quantifying structural similarity through the ratio of successfully aligned amino acids. Figure ~\ref{fig:flapp_dist} illustrates the impact of varying FLAPP score thresholds on the number of remaining samples. To strike a balance between removing highly similar pockets and retaining an adequate volume of training data, we selected thresholds of 0.6 and 0.9, resulting in two distinct datasets, as detailed in Table ~\ref{tab:threshold_data_sizes}. These datasets were subsequently divided into training and validation sets in a 9:1 ratio through random sampling.

\begin{table}[h]
\centering
\caption{Data sizes for different thresholds}
\begin{tabular}{c|c}
\toprule
Threshold (FLAPP Score) & Data Size (PDBbind) \\ 
\noalign{\vskip 2pt} 
\hline
\noalign{\vskip 2pt} 
0.6 & 12,344 \\ 
\noalign{\vskip 2pt} 
\hline
\noalign{\vskip 2pt} 
 0.9 & 17,519 \\
\bottomrule
\end{tabular}
\label{tab:threshold_data_sizes}
\end{table}

\begin{figure}[htbp]
\centering
\includegraphics[width=0.8\textwidth]{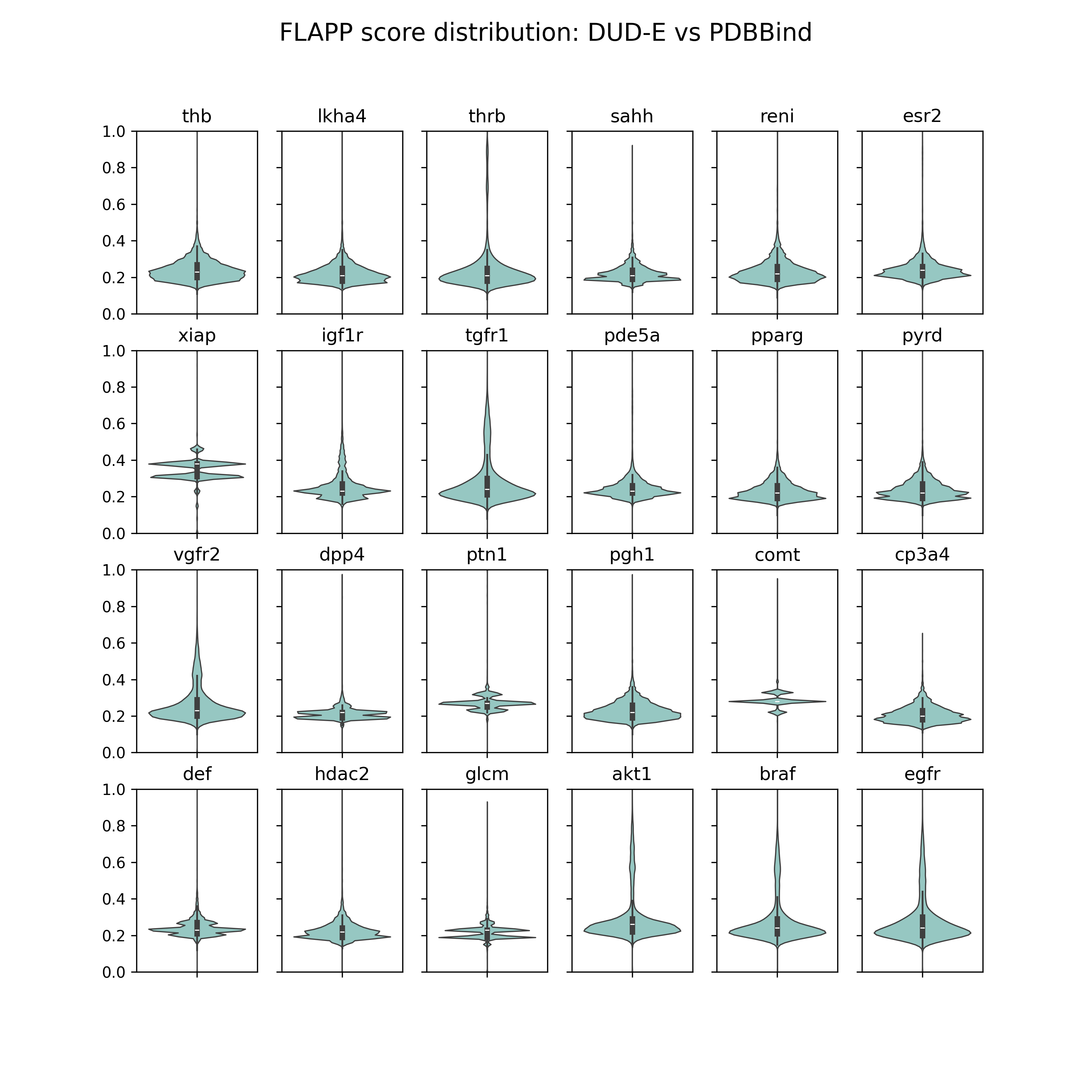}
\caption{FLAPP score distribution. Each figure is the distribution of FLAPP score between a DUD-E target and all pdbbind pockets.}
\label{fig:flapp_dist_per_case}
\end{figure}

Figure ~\ref{fig:flapp_dist_per_case} illustrates the FLAPP similarity scores between selected targets from the DUD-E dataset and protein pockets from the PDBbind database. The majority of cases have scores clustered in the range of 0.2 to 0.4.

\section{Test Set Details}

We constructed new test sets utilizing data from the well-established virtual screening benchmarks DUD-E and LIT-PCBA. Following the removal of erroneous records, we curated 101 test data points from DUD-E and 15 from LIT-PCBA. These test data cover various categories of protein targets, such as G-Protein Coupled Receptors (GPCRs), kinases, and nuclear receptors. This diversity enables a comprehensive assessment of the model's performance across different protein types. The classification and quantity of these data points are provided in Table ~\ref{tab:dude}.

In DUD-E, each target in our test set contains an average of 224.4 active compounds and 50 decoys per active. For LIT-PCBA, each target includes an average of 503.33 active compounds and 176,268.13 decoys. Consistent with our training and validation sets, we defined pockets for SBDD inputs by selecting regions within a 10 Å radius from the reference ligand. This systematic approach ensures a robust and comprehensive evaluation of the model's capabilities across diverse protein-ligand interactions.

\begin{table}[ht]
\centering
\caption{Distribution of protein target categories in our test set}
\begin{tabular}{c|c|c}
\toprule
Target Categories & DUD-E&  LIT-PCBA  \\  \hline
Kinase                     & 26      &      2            \\
Protease                   & 15      &     0             \\
Nuclear Receptor           & 11      &     4              \\
GPCR                       & 4       &   2                \\
Miscellaneous              & 5       &     0             \\
Ion Channel                & 2       &     0             \\
Cytochrome P450            & 2       &     0             \\
Other Enzymes              & 36      &    7               \\
\bottomrule
\end{tabular}
\label{tab:dude}
\end{table}

\section{Model Training Details}

We employed a new dataset to train and test across five distinct SBDD baselines. All models were trained on a single NVIDIA A100 80GB GPU. The training durations were as follows: MolCRAFT required approximately 30 hours, TargetDiff took around 48 hours, Pocket2Mol also took about 48 hours, AR's main model and frontier model each required 48 hours, and LiGAN training took approximately 30 hours.

The code for models used in our benchmark is from their official repositories: 

LiGAN: \href{https://github.com/mattragoza/LiGAN}{https://github.com/mattragoza/LiGAN}

AR: \href{https://github.com/luost26/3D-Generative-SBDD}{https://github.com/luost26/3D-Generative-SBDD}

Pocket2Mol: \href{https://github.com/pengxingang/Pocket2Mol}{https://github.com/pengxingang/Pocket2Mol}

TargetDiff: \href{https://github.com/guanjq/targetdiff}{https://github.com/guanjq/targetdiff}

MolCRAFT: \href{https://github.com/AlgoMole/MolCRAFT}{https://github.com/AlgoMole/MolCRAFT}

We did some modifications to fit our dataset. 
Modified Code for each model can be found at \href{https://github.com/bowen-gao/sbdd_practical_evaluation}{https://github.com/bowen-gao/sbdd\_practical\_evaluation}

\section{Model Sampling Details}

For the 101 + 15  targets in our test set, we sampled 20 small molecules per target using each model. For the autoregressive-based models, AR and TargetDiff, which initially tend to generate smaller molecules, we first sampled 100 molecules and then randomly selected 20 from this set to ensure uniformity in molecule size.

All the sampling results can be found at \\
\href{https://huggingface.co/datasets/bgao95/Practical_SBDD}{https://huggingface.co/datasets/bgao95/Practical\_SBDD}

\section{Evaluation}

For Morgan Fingerprint, we use radius = 2, length of bit vector = 1024.

For E3FP Fingerprint, we use length of bit vector = 1024, radius multiplier = 1.5.

For Uni-Mol molecular encoder, we use the trained weights provided in \href{https://github.com/dptech-corp/Uni-Mol}{https://github.com/dptech-corp/Uni-Mol}

For DrugCLIP related models, we use the trained weights provided in \href{https://github.com/bowen-gao/DrugCLIP}{https://github.com/bowen-gao/DrugCLIP}



\section{Additional Benchmark Results}

\subsection{Distribution Plot for Similarity to Known Drugs in FDA approved list} \label{sec: appendix fda sim}
\begin{figure}[h!]
    \centering
    \begin{subfigure}[b]{0.4\textwidth}
        \includegraphics[width=\textwidth]{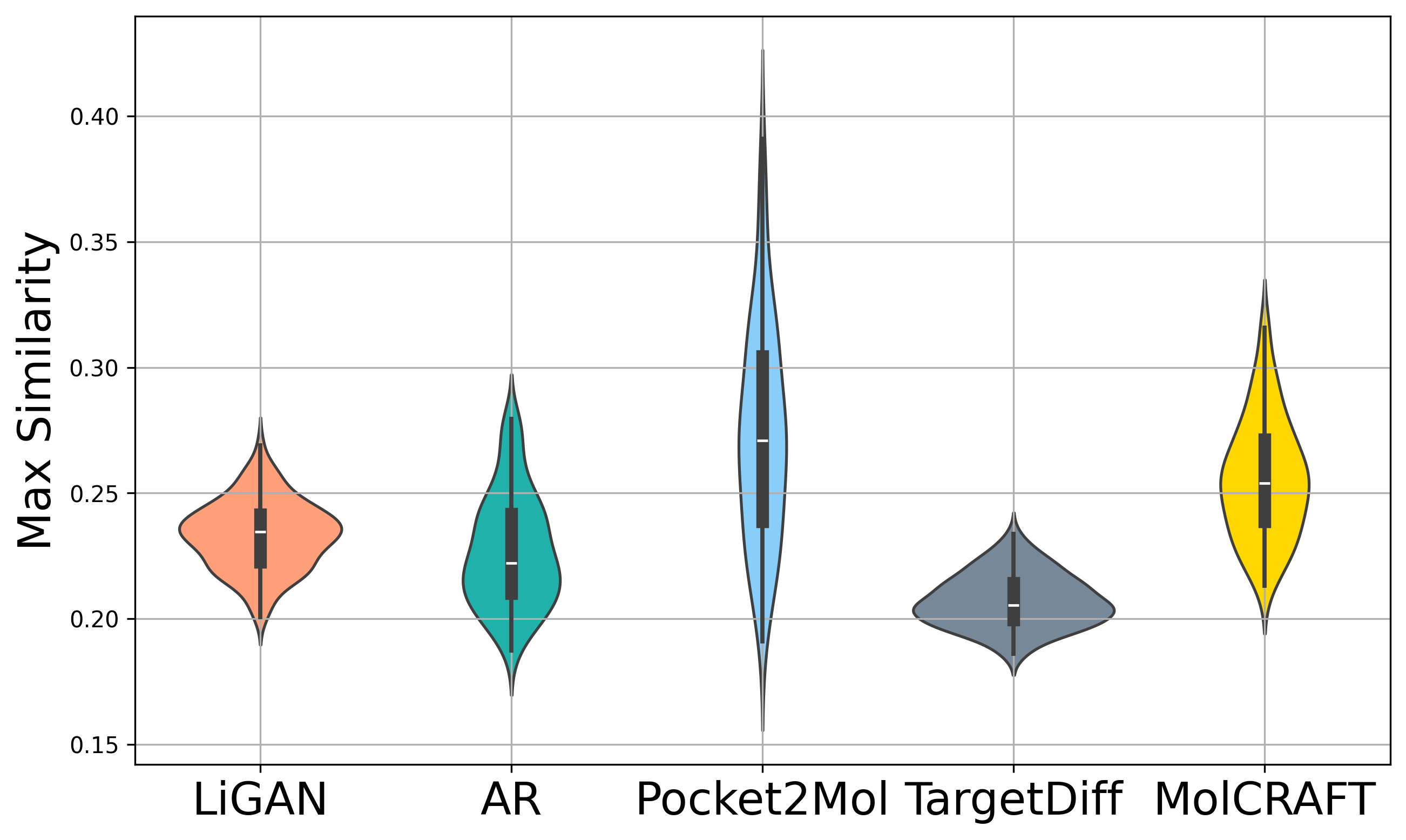}
        \caption{Morgan Fingerprints}
        \label{fig:image1}
    \end{subfigure}
    \hfill 
    \begin{subfigure}[b]{0.4\textwidth}
        \includegraphics[width=\textwidth]{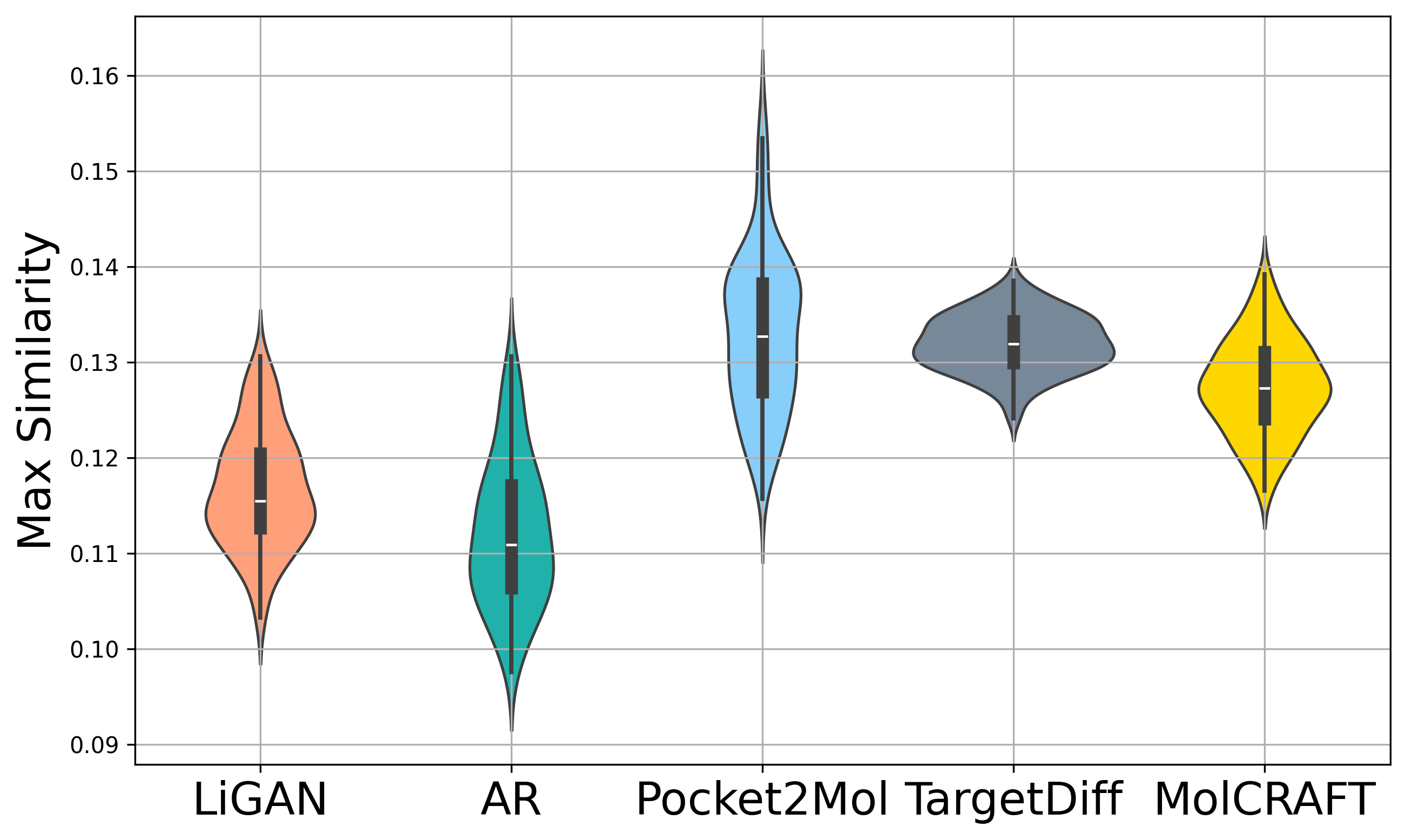}
        \caption{ECFP3 Fingerprints}
        \label{fig:image2}
    \end{subfigure}
    
    \begin{subfigure}[b]{0.4\textwidth}
        \includegraphics[width=\textwidth]{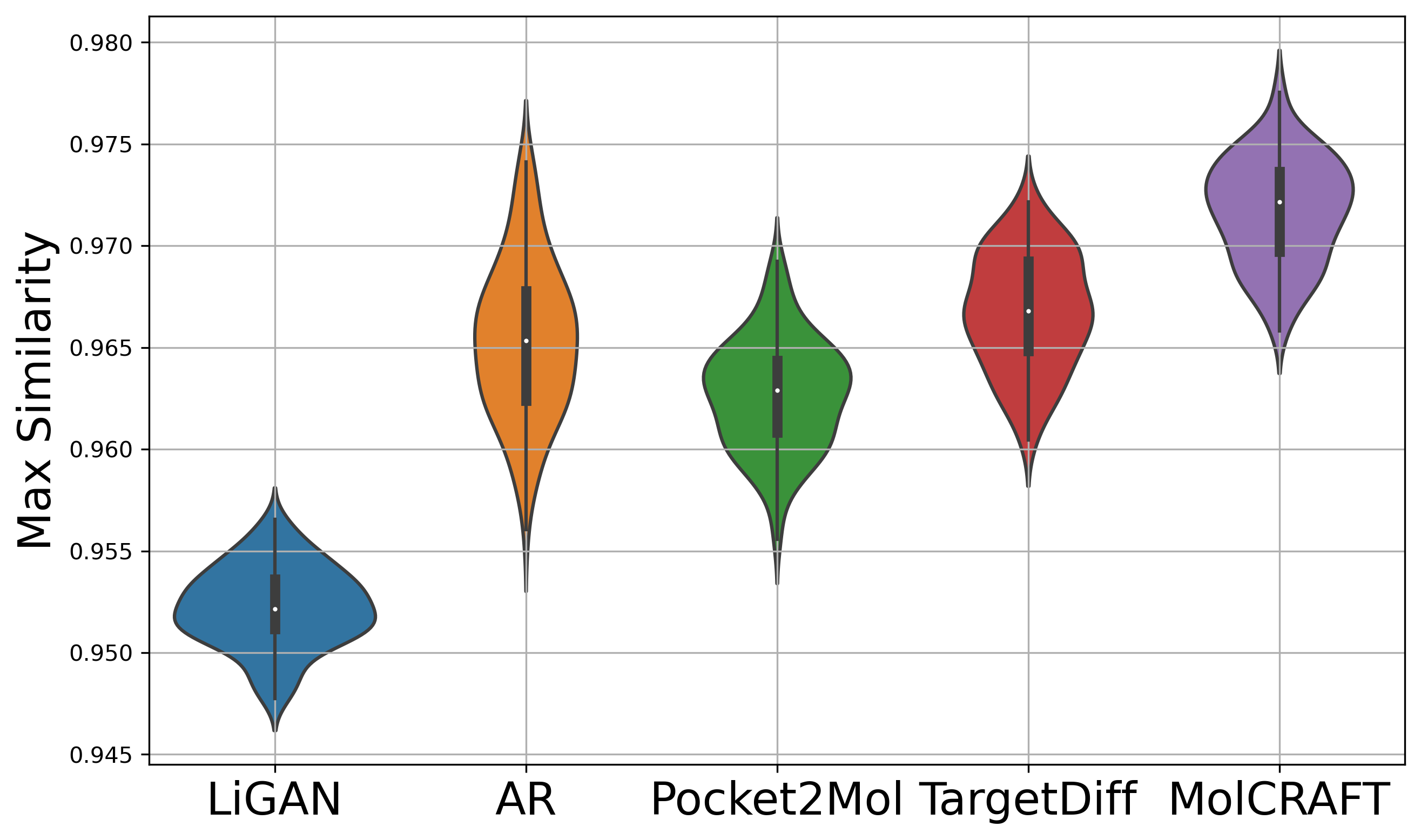}
        \caption{Uni-Mol Molecular Encoder}
        \label{fig:image3}
    \end{subfigure}
    \hfill 
    \begin{subfigure}[b]{0.4\textwidth}
        \includegraphics[width=\textwidth]{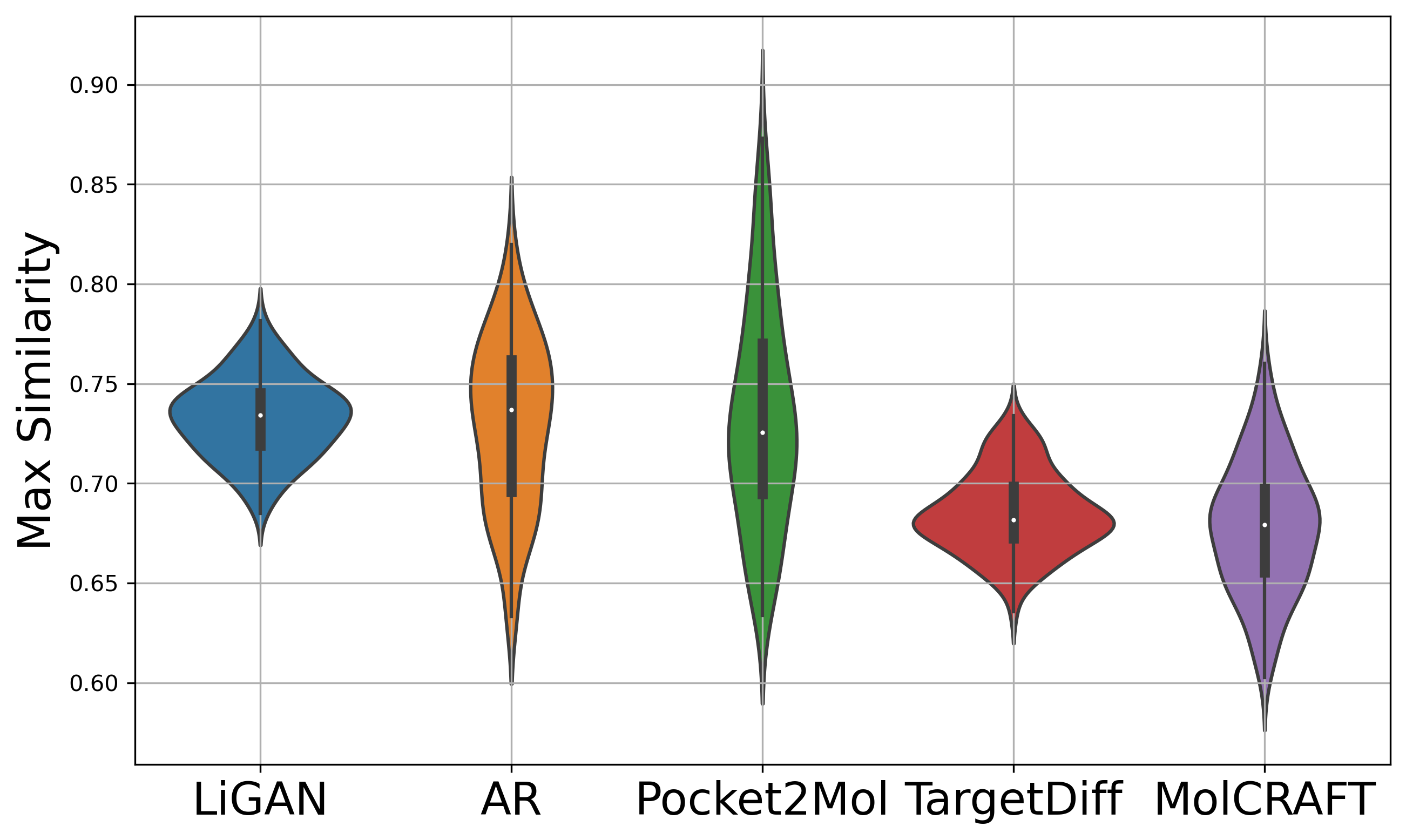}
        \caption{DrugCLIP Molecular Encoder}
        \label{fig:image4}
    \end{subfigure}
    
    \caption{Distribution plots for max similarity to all drugs in FDA approves list on all targets in DUD-E with different models.}
    \label{fig:dist_fda}
\end{figure}

In Figure \ref{fig:dist_fda}

\subsection{Distirbution Plot for Similarity to Known Actives} \label{sec: appendix actives sim}
\begin{figure}[h!]
    \centering
    \begin{subfigure}[b]{0.42\textwidth}
        \includegraphics[width=\textwidth]{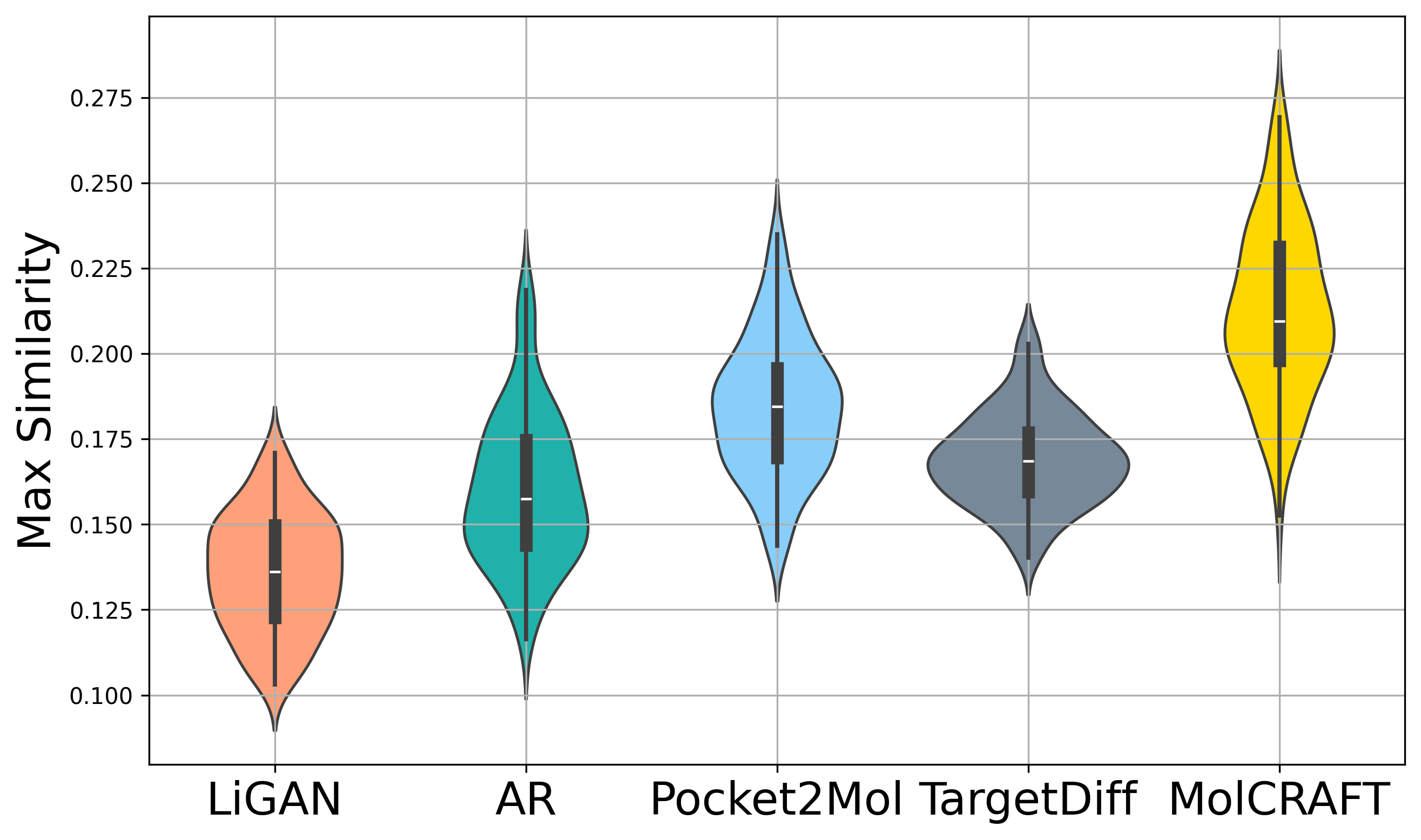}
        \caption{Morgan Fingerprints}
        \label{fig:image1}
    \end{subfigure}
    \hfill 
    \begin{subfigure}[b]{0.42\textwidth}
        \includegraphics[width=\textwidth]{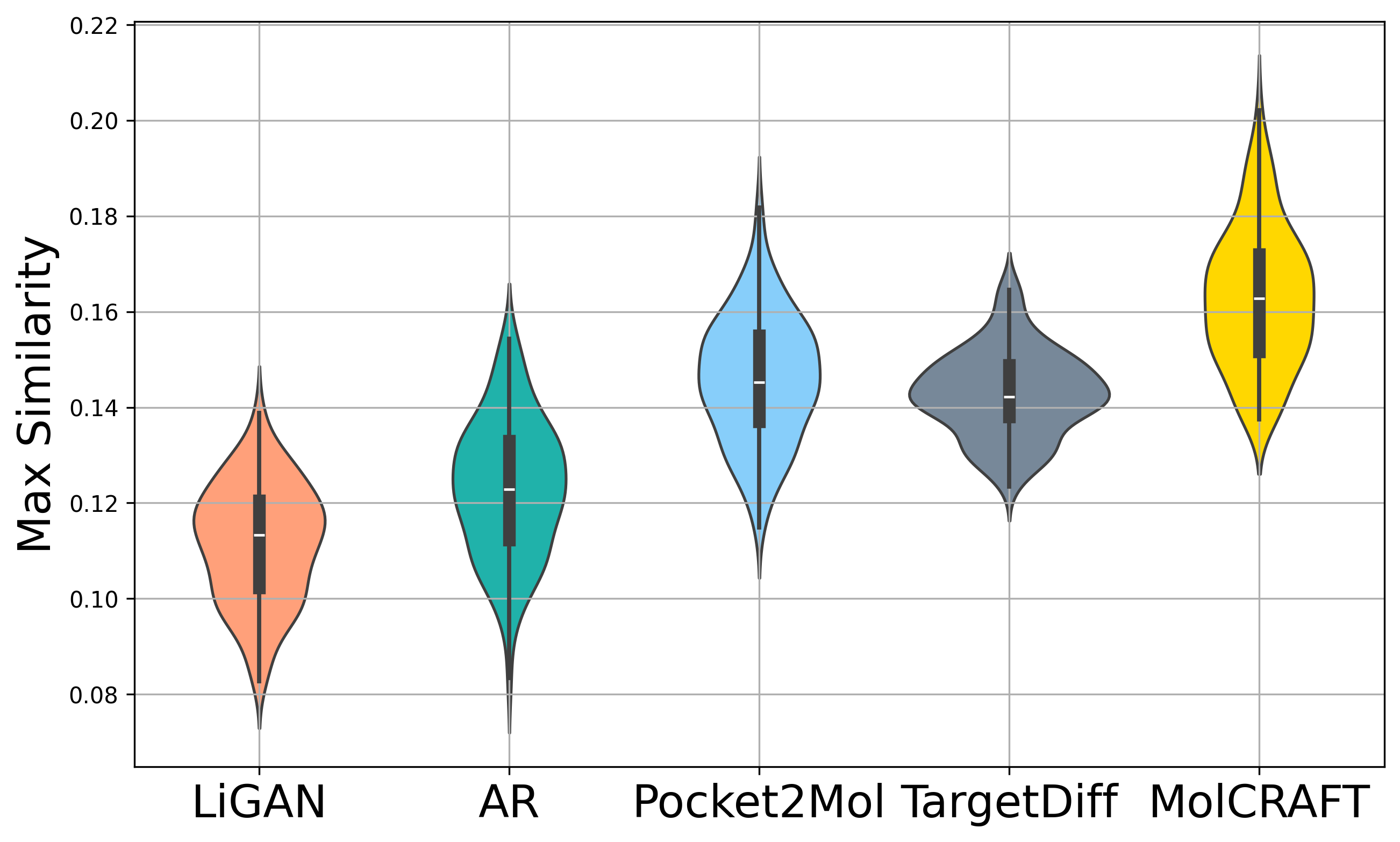}
        \caption{ECFP3 Fingerprints}
        \label{fig:image2}
    \end{subfigure}
    
    \begin{subfigure}[b]{0.42\textwidth}
        \includegraphics[width=\textwidth]{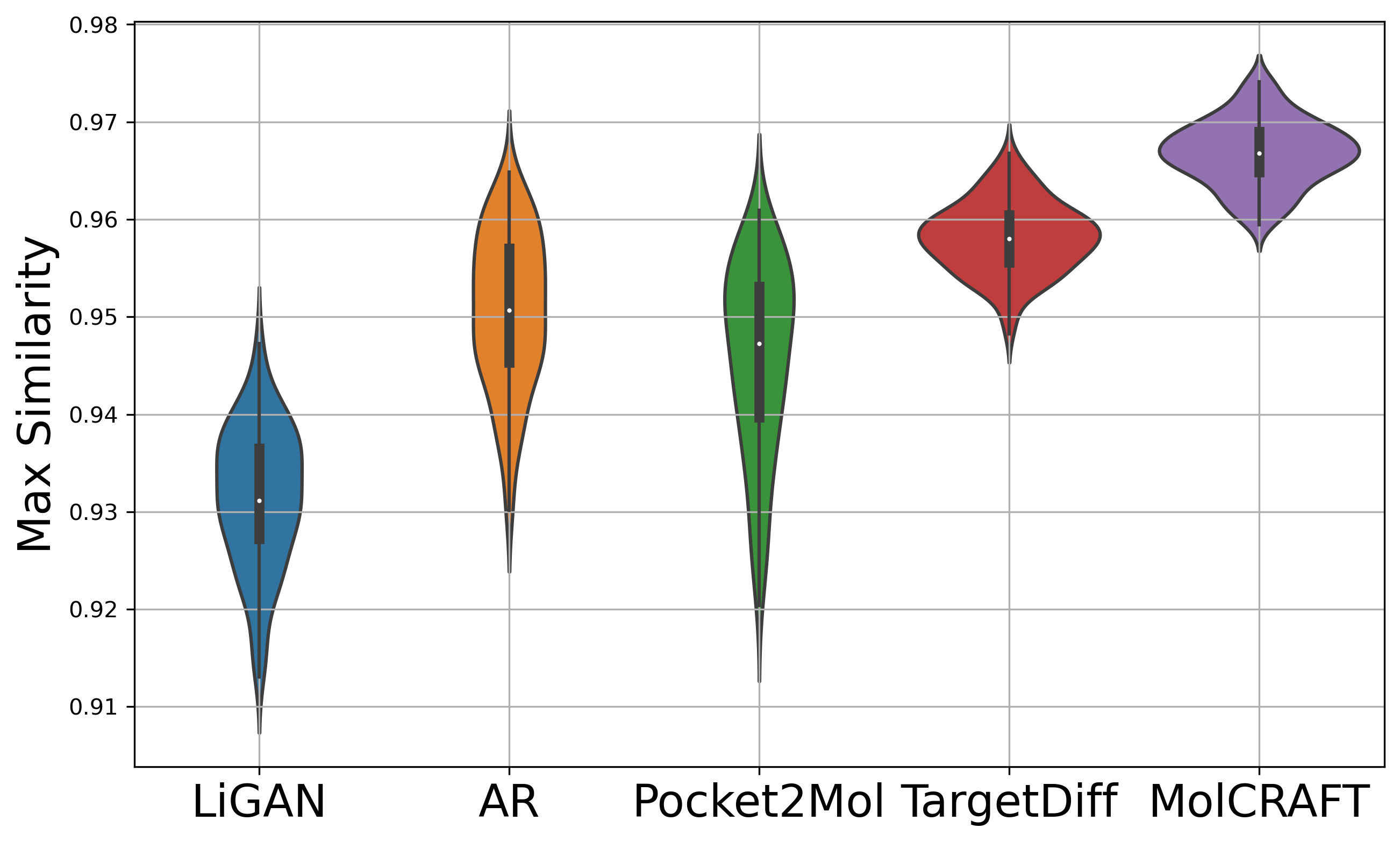}
        \caption{Uni-Mol Molecular Encoder}
        \label{fig:image3}
    \end{subfigure}
    \hfill 
    \begin{subfigure}[b]{0.42\textwidth}
        \includegraphics[width=\textwidth]{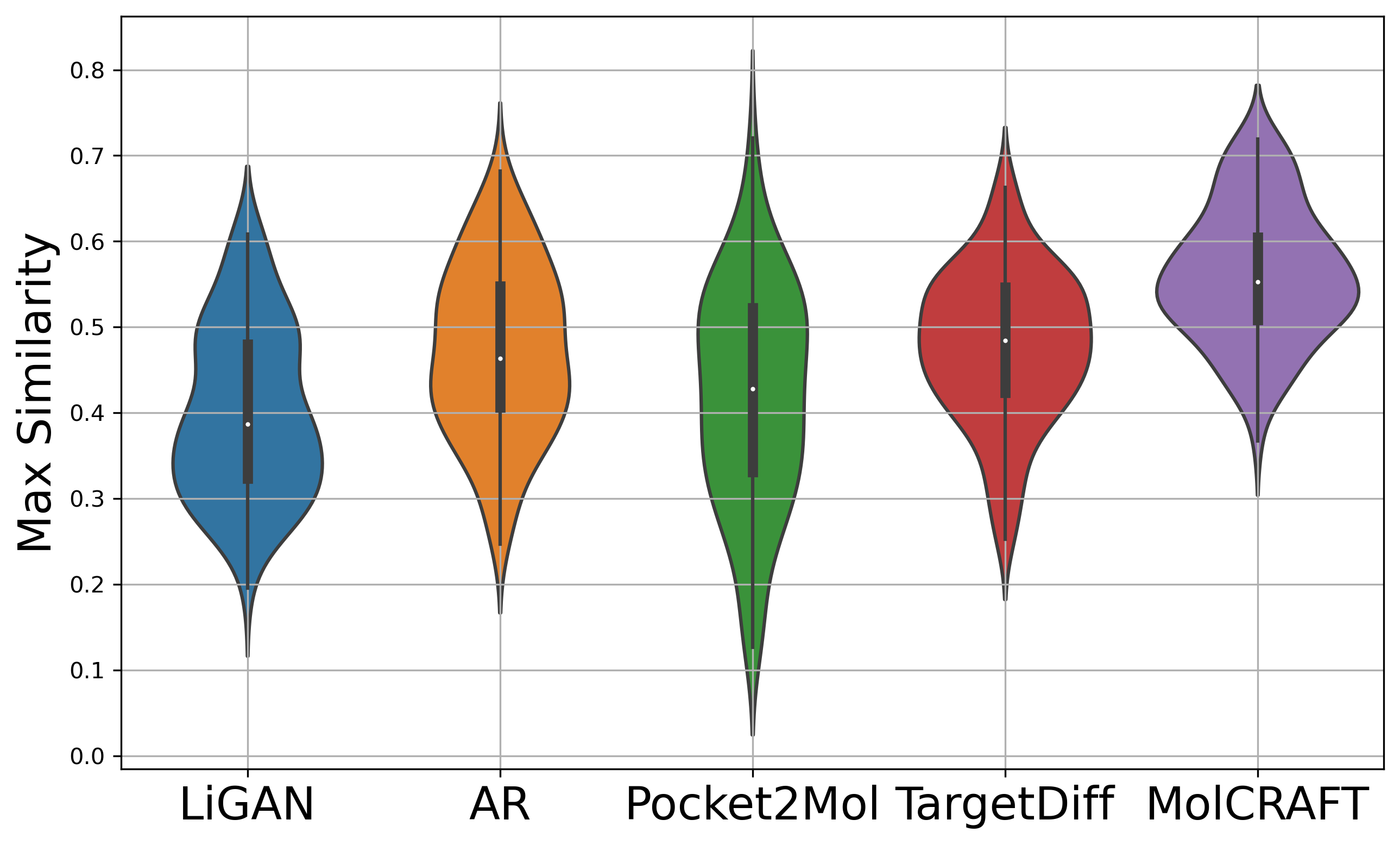}
        \caption{DrugCLIP Molecular Encoder}
        \label{fig:image4}
    \end{subfigure}
    
    \caption{Distribution plots for max similarity to known actives on all targets in DUD-E with different models.}
    \label{fig:dist_active}
\end{figure}

In Figure \ref{fig:dist_active}.

\subsection{Distirbution Plot for Virtual Screening} \label{sec:appendix vs}
\begin{figure}[h]
    \centering
    \begin{subfigure}[b]{0.42\textwidth}
        \includegraphics[width=\textwidth]{figures/violin_plot_ef_morgan.png}
        \caption{Morgan Fingerprints}
        \label{fig:image1}
    \end{subfigure}
    \hfill 
    \begin{subfigure}[b]{0.42\textwidth}
        \includegraphics[width=\textwidth]{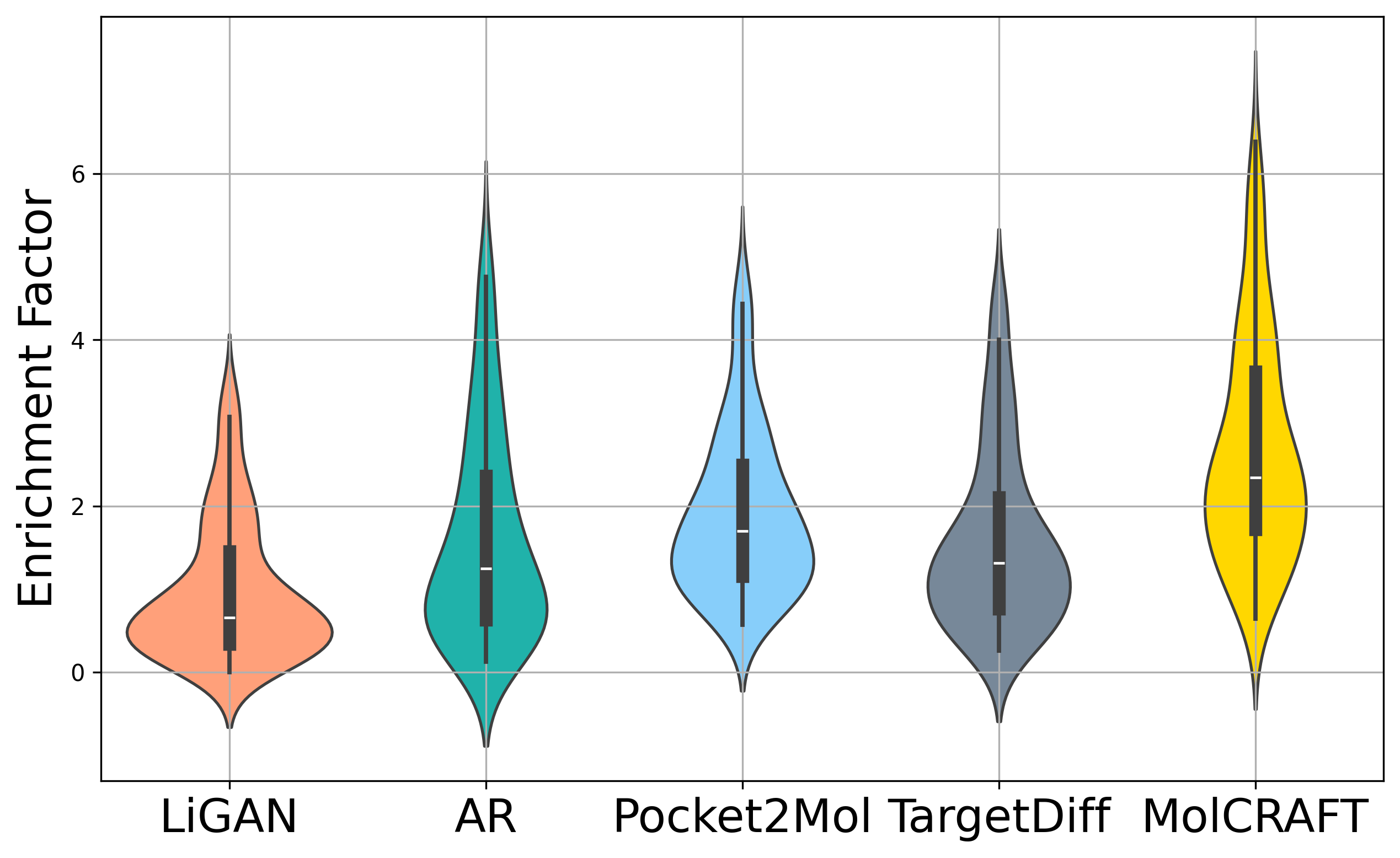}
        \caption{ECFP3 Fingerprints}
        \label{fig:image2}
    \end{subfigure}
    
    \begin{subfigure}[b]{0.42\textwidth}
        \includegraphics[width=\textwidth]{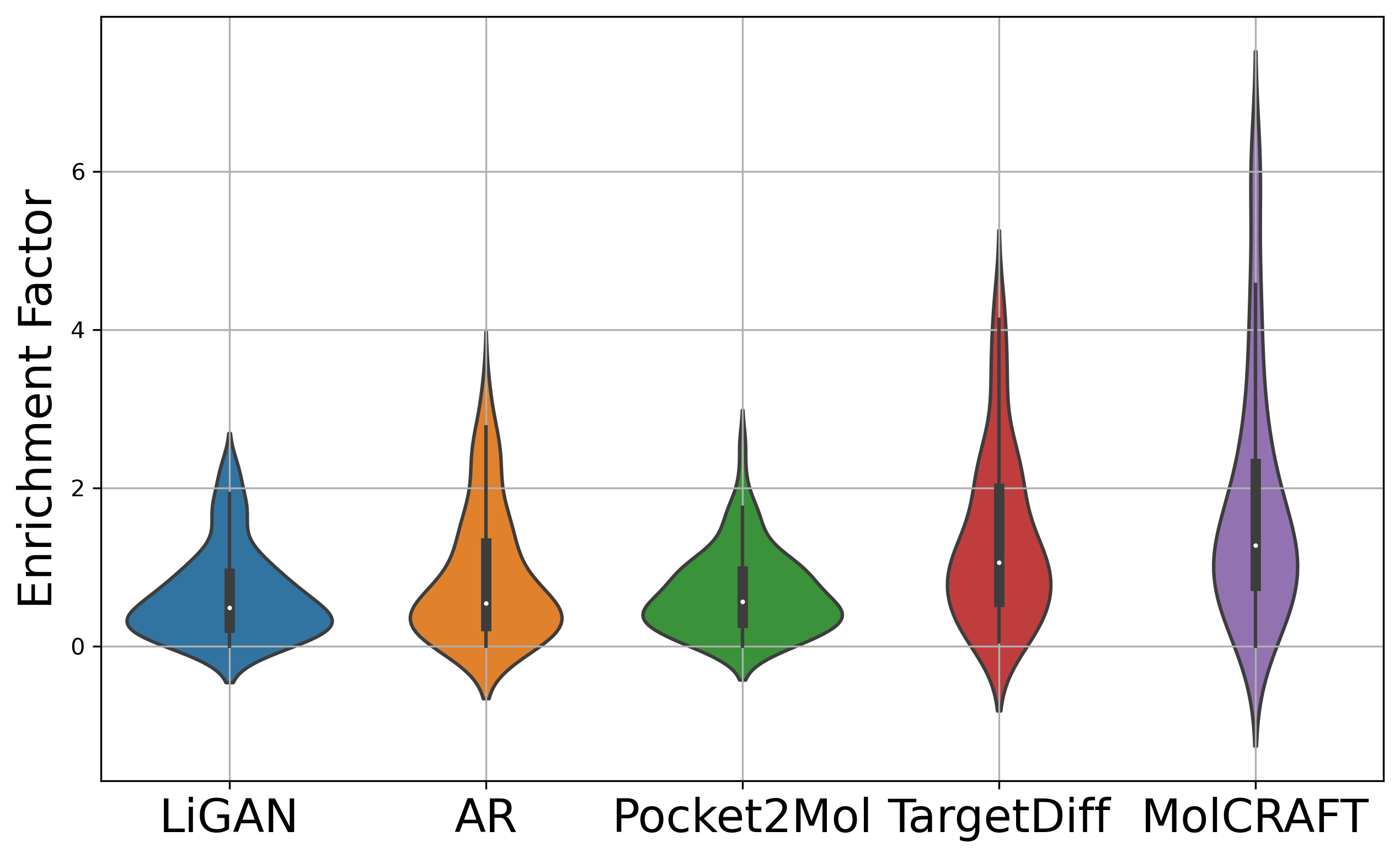}
        \caption{Uni-Mol Molecular Encoder}
        \label{fig:image3}
    \end{subfigure}
    \hfill 
    \begin{subfigure}[b]{0.42\textwidth}
        \includegraphics[width=\textwidth]{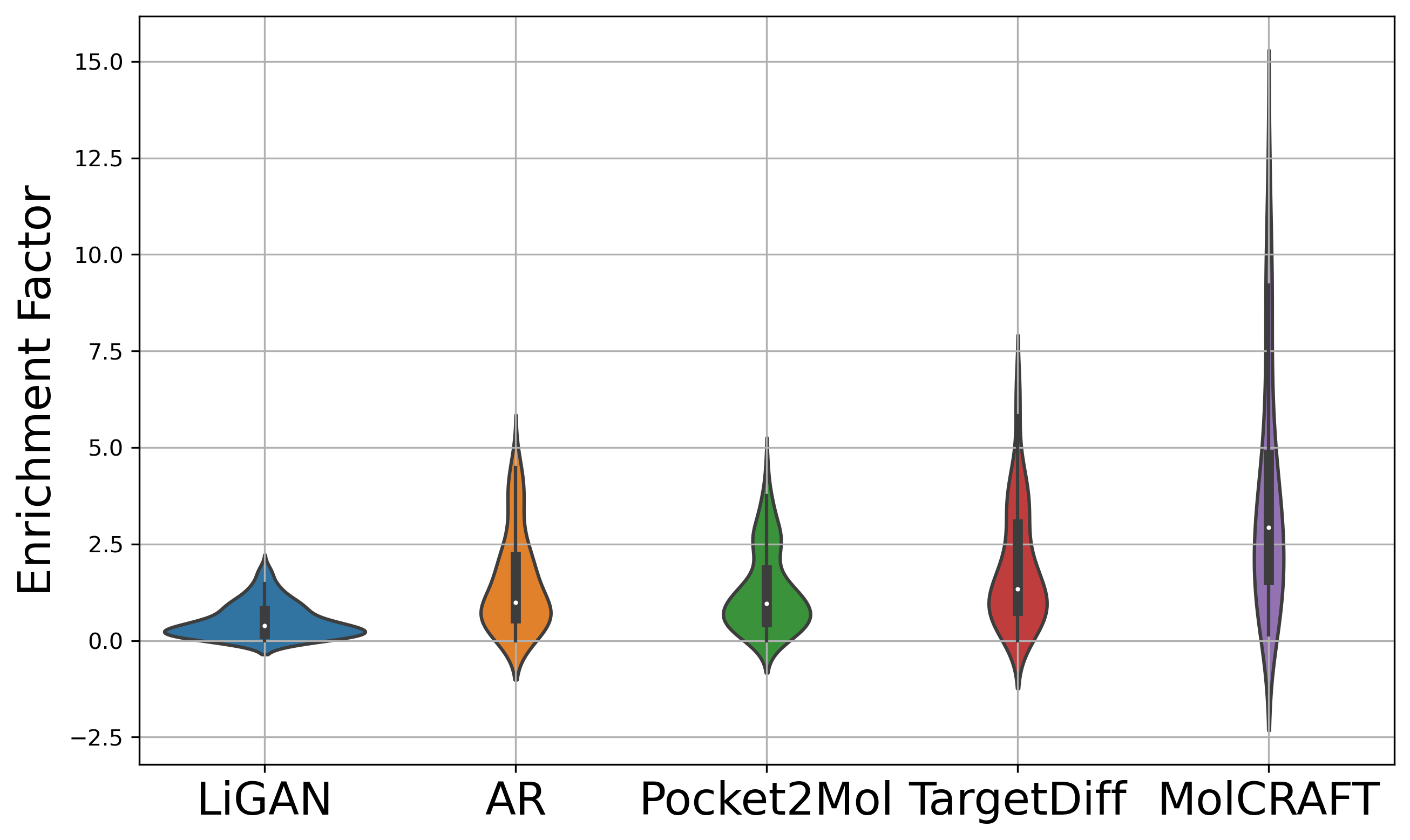}
        \caption{DrugCLIP Molecular Encoder}
        \label{fig:image4}
    \end{subfigure}
    
    \caption{Distribution plots for Virtual Screening results on all targets in DUD-E with different models.}
    \label{fig:dist_vs}
\end{figure}

In Figure \ref{fig:dist_vs}

\subsection{Comparison of PDBbind and CrossDocked dataset.} \label{sec: appendix compare}

\begin{table}[h]
\centering
\caption{Glide Docking score and Delta Score of the pocket with reference ligand on PDBbind and CrossDocked dataset.}
\begin{tabular}{c|cc}
\toprule

            &Glide Docking  &Delta Score    \\ 
            \noalign{\vskip 2pt} 
            \hline
\noalign{\vskip 2pt} 
CrossDocked  & -6.37               & 1.05  \\ 
\noalign{\vskip 2pt} 
\hline
\noalign{\vskip 2pt} 
 PDBBind    & -7.19               & 1.88     \\ 
\bottomrule
\end{tabular}
\label{tab:training_set_metrics}
\end{table}

\section{Limitations}

Although we strive to include a diverse range of SBDD models in our benchmark, it does not encompass all existing models, which is a limitation. Additionally, the set of targets used as our test set is also limited.

\section{Potential Negative Impact}

There is no apparent potential for negative social impact.

\end{document}